\documentclass[%
reprint,
superscriptaddress,
preprintnumbers,
nofootinbib,
amsmath,amssymb,
aps,
prd,
]{revtex4-2}

\usepackage{graphicx}
\usepackage{dcolumn}
\usepackage{bm}
\usepackage[colorlinks=true,urlcolor=purple,anchorcolor=blue,citecolor=blue,filecolor=blue,linkcolor=red,menucolor=blue]{hyperref}

\usepackage[dvipsnames,svgnames]{xcolor}

\usepackage{amsmath,amssymb,graphicx,makeidx,url,wrapfig,epsfig,enumitem}
\usepackage[tight,footnotesize]{subfigure}

\usepackage{color}
\usepackage{cleveref}

\def\m0{m_0}

\def\spose#1{\hbox to 0pt{#1\hss}}
\def\ltapprox{\mathrel{\spose{\lower 3pt\hbox{$\mathchar"218$}}
 \raise 2.0pt\hbox{$\mathchar"13C$}}}
\def\gtapprox{\mathrel{\spose{\lower 3pt\hbox{$\mathchar"218$}}
 \raise 2.0pt\hbox{$\mathchar"13E$}}}
\def\inapprox{\mathrel{\spose{\lower 3pt\hbox{$\mathchar"218$}}
 \raise 2.0pt\hbox{$\mathchar"232$}}}

\newcommand{\mnb}[0]{\ensuremath{m_{n,0}}}
\newcommand{\mlb}[0]{\ensuremath{m_{l,0}}}

\newcommand{\mdm}[0]{\ensuremath{M_{B_d}}}
\newcommand{\tr}[0]{\ensuremath{{\rm Tr}}}
\newcommand{\hc}[0]{\ensuremath{{\rm h.c.}}}
\newcommand{\meq}[0]{\ensuremath{m_{\rm eq}}}

\begin{document}

\preprint{LLNL-JRNL-2001590, FERMILAB-PUB-24- 0954-T}

\title{Hyper Stealth Dark Matter and Long-Lived Particles}

\author{George T. Fleming}
\affiliation{Theoretical Physics Division, Fermilab, Batavia, IL 60510, USA}
\author{Graham D. Kribs}
\affiliation{Institute for Fundamental Science and Department of Physics, \\
  University of Oregon, Eugene,
  Eugene, OR 97403 USA}
\author{Ethan T. Neil}
\affiliation{Department of Physics, University of Colorado, Boulder,
  CO 80309, USA}
\author{David Schaich}
\affiliation{Department of Mathematical Sciences, University of
  Liverpool, Liverpool L69 7ZL, United Kingdom}
\author{Pavlos M. Vranas}
\email{vranas2@llnl.gov}
\affiliation{Physical and Life Sciences, Lawrence Livermore National
  Laboratory, Livermore, CA 94550, USA}
\affiliation{Nuclear Science Division, Lawrence Berkeley National
  Laboratory, Berkeley, CA 94720, USA}


\begin{abstract}
  A new dark matter candidate is proposed that arises as the lightest baryon from a confining 
  $SU(N)$ gauge theory which equilibrates with the Standard Model only through electroweak interactions.  Surprisingly, this candidate can be as light as a few GeV\@.  The lower bound arises from the intersection of two competing requirements:
  i) the equilibration sector of the model must be sufficiently heavy, at least several TeV, to avoid bounds from colliders, and ii) the lightest dark meson (that may be the dark $\eta'$, $\sigma$, or the lightest glueball) has suppressed interactions with the SM, and must decay before BBN\@.  The low energy dark sector consists of one flavor that is electrically neutral and an \emph{almost} electroweak singlet. 
  The dark matter candidate is the lightest baryon consisting of $N$ of these light flavors leading to a highly suppressed elastic scattering rate with the SM\@.  The equilibration sector consists of vector-like dark quarks that transform under the electroweak group, ensuring that the dark sector can reach thermal equilibrium with the SM in the early Universe.  The lightest dark meson lifetimes vary between $10^{-3} \lesssim c \tau \lesssim 10^7$~meters, providing an outstanding target for LHC production and experimental detection. 
  We delineate the interplay between the lifetime of the light mesons, the suppressed direct detection cross section of the lightest baryon, and the scale of equilibration sector that can be probed at the LHC\@.  
\end{abstract}

\maketitle

{
 \hypersetup{linkcolor=black}
 \tableofcontents
}

\section{Introduction}
\label{sec:intro}

Theories in which dark matter can reach thermal equilibrium with the Standard Model (SM) 
provide an elegant class of well-motivated 
dark matter candidates.  
Broadly there are two classes of theories:
one set that require only the SM interactions
in order to equilibrate the dark sector with
the SM, and another set 
that rely on new (dark sector) mediators
that could have a wide range in mass.  
Extensive efforts
to search for new mediators through a variety of experimental and observational probes are underway, from beam dumps and astrophysical probes of light mediators 
\cite{Alexander:2016aln,Lanfranchi:2020crw}
to the LHC to probe TeV mediators
\cite{Albert:2017onk,Blanco:2019hah}.  But, none of these probes has uncovered evidence
for a new mediator.  This motivates  
reconsidering the more economical class of dark matter theories where the only interactions between the dark sector and the SM are the mediators of SM interactions themselves.

For dark matter candidates that are elementary particles, much of the theory space has already been constrained by a combination of direct detection experiments and collider constraints. 
A small number of weakly interacting massive particle (WIMP) candidates remain, such as the wino and Higgsino in supersymmetric theories \cite{Hisano:2005ec,Hisano:2010fy,Krall:2017xij} and related 
candidates in electroweak multiplets \cite{Cirelli:2005uq}.
Even setting aside thermal abundance calculations, these candidates cannot be lighter than a few hundred GeV,
because otherwise their electrically charged partners
(e.g., charged wino or charged Higgsino) would have been visible in LHC searches \cite{ATLAS:2021yqv,ATLAS:2022rme,CMS:2022sfi}.\footnote{While there are caveats to some LHC search bounds, there are at least robust bounds on new electrically charged particles from LEP II \cite{LEPchargebound1,LEPchargebound2}, roughly $\gtrsim 100$~GeV\@.} 
Composite dark matter candidates 
involving electrically charged 
constituents 
\cite{Kribs:2009fy,Appelquist:2015yfa,Antipin:2015xia,Mitridate:2017oky,Asadi:2024bbq}
also have significant bounds 
arising from the meson sector, 
e.g.,~\cite{Strassler:2006im,Han:2007ae,Kilic:2009mi,Kilic:2010et,Fok:2011yc,Kribs:2018ilo,Beauchesne:2018myj,Cheng:2021kjg}
that can also contain electrically charged particles.
For example, Stealth Dark Matter \cite{Appelquist:2015yfa,Appelquist:2015zfa} is composed of electrically charged dark fermion constituents, and yet has suppressed interactions with the SM where the leading
interaction arises through the electromagnetic polarizability operator \cite{Appelquist:2015zfa}. But even in this model, the lower bound on dark matter is at least 
a few hundred GeV,
based on a combination of collider bounds 
on dark mesons \cite{Kribs:2018oad,Kribs:2018ilo} (e.g., for a recent ATLAS study see \cite{ATLAS:2024xbu})
combined with direct detection constraints.

The purpose of this paper is to investigate strongly coupled theories that can evade the collider constraints, yielding composite dark matter much lighter than previously considered.  We restrict our consideration to theories in which i) dark matter achieves thermal equilibrium with the SM, and ii) the only mediators to the dark sector are the SM interactions:  the Higgs boson and the SM gauge bosons.  The first assumption implies that a relic abundance is generated with at least some connection to the Standard Model abundance due to the thermal contact.
The second assumption simplifies the theory space, and of course is consistent with current experimental and observational nonobservation of new mediators. 

Our focus is on dark matter candidates that are composite baryonic states made from dark fermions that have very small but nonzero couplings to the SM\@.  Dark baryons are among the best motivated dark matter candidates 
\cite{Kribs:2009fy,Appelquist:2015yfa,Antipin:2015xia,Mitridate:2017oky}
since a confining $SU(N)$ gauge theory with $N \ge 3$ provides an automatic accidental symmetry, dark baryon number, that stabilizes dark matter (at least up to dimension-$2 N$ interactions), just like the proton in the SM\@.  This is in contrast to dark mesons, where there is no automatic conserved quantum number.  (There are specific exceptions, e.g.~\cite{Bai:2010qg,Buckley:2012ky}.)  

One might think that it is trivial to obtain arbitrarily light composite baryonic dark matter by simply taking the dark confinement scale (and the dark quark mass scale) arbitrarily small.  
Restricting to theories that have baryonic states, the minimal number of mesons arises in theories with just  one (Dirac) fermion flavor.  This has been explored before for $SU(2)$ in \cite{Francis:2018xjd} and $SU(N)$ in \cite{Morrison:2020yeg}.  There are a few possible candidates for the lightest meson in these one-flavor theories, depending on the relative hierarchy between the dark fermion mass (of the one flavor) and the dark confinement scale.  One distinct possibility is  
the lightest (pseudo)scalar meson, the dark analogue of the $\eta'$ of QCD, which we'll denote as $\eta'_d$.
The other candidates for light mesons include the $\sigma$
(the $J^{PC} = 0^{++}$ state formed from the $f\bar{f}$ bound state) and the lightest ($J^{PC} = 0^{++}$) glueball.   
The key observation is that confining theories with a
composite baryonic state \emph{must} be accompanied by lighter mesonic states.  
In the \emph{absence of interactions} (other than gravitational interactions) between the dark sector and the SM, the lightest of these mesons and the lightest baryon, denoted by $B_d$, are stable, and this could lead to a multi-component theory of dark matter \cite{Morrison:2020yeg}. 
Of course this assumes there is some mechanism to obtain the correct relic abundance -- e.g., thermal abundance from dark baryon annihilation into dark mesons -- but even this mechanism is not predictive since it requires specifying the initial temperature of the dark sector that is separate from the SM sector \cite{Morrison:2020yeg}. There are additional  restrictions on these states arising from 
baryon-meson interactions that lead to self-interactions among the dark matter states.  If the states are too light, the self-interactions may violate bounds from galaxy cluster mergers (for a review, see \cite{Tulin:2017ara}). 

In this paper, the model we propose has SM interactions between the dark sector and the SM, thereby providing a viable way to thermalize the dark sector with the SM\@.  This is achieved using a ``dark equilibration sector'' that not only equilibrates the SM with the dark sector, but also provides the interactions between the dark sector and SM that permit all of the dark mesons to decay.
The dark sector consists of a confining $SU(N)$ with 
$N \ge 3$ with one flavor of dark quark that is very light and neutral under the SM gauge group.  
The theory is also accompanied by 
several heavy fermion flavors with full-strength electroweak interactions that serve as the thermal equilibration sector.  Interestingly, suppressed Higgs interactions between the light and heavy flavors causes tree-level mixing
between the one neutral light flavor and 
one flavor in the equilibration sector.  The mixing implies the light flavor \emph{mass
eigenstate} is \emph{almost}, but not exactly, an electroweak singlet.
This serves two critical purposes:  
first, the lightest baryon -- the dark matter candidate in our paper -- will have very small elastic scattering off SM particles, which can easily be well below the current direct detection bounds.
This is the origin of the name for this 
model framework: Hyper Stealth Dark Matter (HSDM). 
The second purpose is that the lightest 
parity-odd meson ($\eta_d'$), that is composed
of a bound state of the light, almost-electroweak singlet flavor, can decay through the suppressed
interactions with the SM\@.  The interactions
are necessarily suppressed because the 
equilibration sector must be above at least
a few TeV, or otherwise the LHC should have seen
evidence of this electroweak-charged meson sector.
The hierarchy between the light 
almost-electroweak singlet flavor 
and the equilibration sector flavors is what
leads to the light mesons being
necessarily long-lived throughout the
parameter space where the dark matter is 
relatively light.  HSDM thus provides a 
fantastic motivation for continued studies of
long-lived particles both at the LHC \cite{Alimena:2019zri} and 
future dedicated experiments such as 
FASER \cite{FASER:2018eoc} and MATHUSLA \cite{MATHUSLA:2018bqv,MATHUSLA:2020uve}.

In the following, we motivate HSDM, define it and its parameters,
present the final results for the case of non-heavy fermions where the
mesonic states are mainly fermionic, and discuss its physical
signatures. Also, the details of these calculations are presented
along with the definition and calculation of the resulting one-flavor
low energy effective field theory, the robustness of HSDM for
different values of its parameters, the interactions with the Higgs,
and other bounds arising from the dark sigma meson and glueball
states, the latter of which become important when the lightest
fermions become very heavy compared to the dark confinement scale. For
some other interesting works see \cite{Farakos:2025ukq,
Yamanaka:2019aeq, Yamanaka:2019yek, Zhang:2021orr, Park:2017rfb}.

\section{Hyper Stealth Dark Matter}
\label{sec:hsdm}

HSDM is a theory of dark matter consisting of two distinct sectors in one theory:  a light dark matter sector, and a heavy dark equilibration sector.  These sectors have distinct roles in the theory.  The dark matter sector can be formulated as a low energy effective theory that consists of one light (Dirac)
flavor transforming under a confining $SU(N)$ gauge theory with higher dimensional interactions with the SM\@.  The higher dimensional interactions are generated by integrating out the equilibration sector, that consists of several heavy (Dirac) flavors that transform under the electroweak interactions of the SM\@.  In addition, a major role of the
equilibration sector is to ensure the dark sector is able to come into thermal equilibrium with the SM (so long as the reheat temperature is above the scale of the equilibration sector), providing at least one mechanism -- thermal freeze out -- that
can generate the dark matter relic abundance.  The suppressed SM interactions introduced as a result of this structure open pathways for experimental detection of the dark sector.  But, as we will see, the scale of the equilibration sector is constrained by the phenomenological constraints within the dark matter sector itself, and so there is a tight relationship between the two sectors.

\subsection{Dark Matter Sector \label{subsec:dm_eft}}

The dark matter sector interaction Lagrangian is given in Eq. \ref{eq:Lint_EW_symmetric}. This is the most general Lagrangian
for HSDM type of theories with UV completions of the type described in this section and in section \ref{sec:equilibration}.
It  consists of a single light Dirac fermion $\Psi_n$, which transforms in the fundamental representation of an SU$(N_D)$ dark gauge interaction.  This yields as a dark matter candidate the lightest (anti-)baryon $B_d$ ($\bar{B}_d$), which is a bound state of $N_D$ fermions $\Psi_n$ ($\bar{\Psi}_n$) with mass $\mdm$.  Since this is a one-flavor theory, fermion statistics in a quark-model picture requires the baryon to have spin $N_D/2$ \footnote{This argument relies on the ground state being the state of lowest angular momentum in a quark model.  In QCD, this is the case: the spin-3/2 $\Delta^{++}$ resonance is the analogue of our $B_d$, and while a spin-1/2 state does exist, the $\Delta(1620)$, it is about 400 MeV heavier than the ground-state $\Delta^{++}$ with higher spin.  While lattice calculations have shown the lowest-spin baryon to be the lightest in SU(4) quenched theory \cite{LSD:2014obp,Brower:2023rqf} and in large-$N_D$ scaling with two dynamical fermions \cite{DeGrand:2016pur}, further lattice studies will be needed to conclusively verify this picture in the one-flavor case at large $N_D$.}.  Since $\Psi_n$ is an electroweak singlet, so is the dark baryon $B_d$, 
in the absence of mixing with the equilibration sector fermions.  As we will see, both $\Psi_n$ and $B_d$ will have highly suppressed weak interactions and Higgs interactions with the SM through higher-dimensional operators that result from integrating out the equilibration sector.

In addition to the dark matter candidate itself, the dark matter sector also contains a pseudoscalar meson $\eta'_d$, that is the analogue of the $\eta'$ of QCD.  The $\eta'_d$ is also a potential dark matter candidate under certain conditions, as studied in detail in \cite{Morrison:2020yeg} under a ``nightmare scenario'' in which the composite dark sector has no Standard Model interactions.  In this paper, the presence of electroweak operators destabilizes the $\eta'_d$, leaving only the dark baryon $B_d$ as the dark matter, although the lifetime of the $\eta'_d$ will give meaningful constraints on the model parameter space, primarily from big bang nucleosynthesis (BBN), considered in detail in \cref{sec:bbn} below.

Finally, the dark matter sector also has a large number of additional bound states: higher-spin dark mesons, dark glueballs, and additional excited states.  As with the $\eta'_d$, the presence of Standard Model interactions will generally allow them to decay on cosmic timescales.  We will discuss possible constraints from these additional states in \cref{sec:bbn} below.

Turning to interactions, in addition to the strongly-coupled dark gauge interaction, the dark matter sector also has higher-dimensional interactions with the electroweak sector of the SM.  These interactions are suppressed by powers of a heavy mass scale $\Lambda$.  As we will see, the equilibration sector (described below, although more general UV completions may lead to the same effective one-flavor theory) will generate these (and other) operators in dark matter sector.  The possible effective operators couple $\bar{\Psi}_n \Psi_n$ bilinears to electroweak singlet SM operators.  

We focus on a subset of the possible interactions which are most phenomenologically relevant, given the UV completion and the phenomenology to be studied below.  Further discussion of other possible operators is given at the end of this subsection.  We thus consider the following set of interactions:
\begin{eqnarray}
  \mathcal{L}
  &\supset& c_s \frac{\overline{\Psi}_n \Psi_n H^\dagger H}{\Lambda}
  + c_G \frac{\tr[G_{\mu\nu} G^{\mu\nu}] H^\dagger H}{\Lambda^2} \nonumber \\
  & & {}+ c_Z \frac{\overline{\Psi}_n \gamma_\mu \Psi_n (H^\dagger i D^\mu H + \hc)}{\Lambda^2} \nonumber \\
  & & {}+ c_Z' \frac{\overline{\Psi}_n \gamma_\mu \gamma^5 \Psi_n (H^\dagger iD^\mu H + \hc)}{\Lambda^2}
  \label{eq:Lint_EW_symmetric}
\end{eqnarray}
where $D_\mu$ is the standard gauge-covariant derivative including the SU$(2)_L$ and U$(1)_Y$ gauge fields, $\tr[G_{\mu\nu} G^{\mu\nu}]$ is the usual gauge-invariant trace over the squared SU$(N_D)$ field-strength tensor $G_{\mu \nu}^a$. The relative size of the two couplings $c_Z$ and $c_Z'$ will be determined by the details of the UV completion, i.e., the content of the dark equilibration sector.  Note that all of the interactions are invariant under CP conjugation, but the operator with coupling $c_Z'$ is parity violating.  If dark sector confinement occurs below electroweak symmetry breaking, the total mass $m_n$ of the dark fermion is given by
\begin{equation} 
m_n = \mnb + c_s \frac{v^2}{2 \Lambda} \label{eq:mn_simple}
\end{equation}
where $m_{n,0}$ is the bare vector-like mass.

Below the scale of electroweak symmetry breaking, \cref{eq:Lint_EW_symmetric} leads to the following linear interactions with the Higgs and $Z$ boson:
\begin{equation}
\mathcal{L} \supset
   \frac{c_s v}{\Lambda} \overline{\Psi}_n \Psi_n h
    - \frac{v M_Z}{\Lambda^2}
       \overline{\Psi}_n \gamma_\mu \left( c_Z + c_Z' \gamma_5 \right) \Psi_n Z^\mu, 
       \label{eq:nZ_current}
\end{equation}
The dominant interaction for direct detection proceeds through the vector coupling, $c_Z$, since $c_Z'$ will lead to a coupling to the axial current within target nuclei, which is heavily suppressed \cite{Servant:2002hb}.
These interactions imply $B_d$ has 
spin-independent contributions to 
its scattering off nuclei.  The
coupling through the $Z$ boson has strength \cite{Lin:2019uvt}
\begin{equation}
  \langle B_d | j_Z^\mu | B_d \rangle 
  = N_D c_Z \frac{vM_Z}{\Lambda^2} \label{eq:baryon_ME_Z}
\end{equation}
times a vector form factor for the baryon, but at small momentum transfer (relevant for dark matter direct detection) the form factor is simply equal to 1, and is independent of $N_D$ in the large-$N_D$ limit \cite{Flores-Mendieta:1998tfv}.

The $Z$-exchange is the leading interaction of dark matter with the SM\@.  It is highly constrained by direct detection bounds, which will lead to 
significant constraints on the coupling $c_Z v^2 / \Lambda^2$ also depending on the dark matter mass scale $\mdm$.  There are also interactions of $B_d$ with the Higgs boson.  Following \cite{LatticeStrongDynamicsLSD:2014osp}, the contribution from $c_s$ gives rise to a Higgs-dark baryon coupling of the form
\begin{equation}
g_{B_d,h} = \frac{\mdm}{m_n} \frac{c_s v}{2\Lambda} f_n^{(B_d)}
\end{equation}
where $f_n^{(B_d)} \equiv \frac{m_n}{\mdm} \frac{\partial \mdm}{\partial m_n}$ is the ``sigma term'' of the dark nucleon, defined as
\begin{equation} 
f_n^{(B_d)} \equiv \frac{m_n}{\mdm} \langle B_d | \overline{\Psi}_n \Psi_n | B_d \rangle.
\label{eq:sigma_n}
\end{equation}
$f_n^{(B_d)}$ may be computed entirely from the strong dynamics, e.g., using a lattice simulation. 

In principle, there is an additional contribution to the baryon-Higgs coupling from the operator $c_G$ in \cref{eq:Lint_EW_symmetric}, which can be thought of as arising from the Higgs coupling to heavy fermions in the equilibration sector.  The contribution from this operator is estimated in \cref{app:trace_anomaly}; we find it to be generally negligible compared to the coupling due to valence $\Psi_n$ fermions in the baryon.

Finally, the electroweak interaction $c_Z'$ will mediate the decay of the $\eta'_d$ dark meson into pairs of SM fermions.  Below the dark confinement scale $\Lambda_d$, we can match on to a chiral effective theory which includes the $\eta'_d$ explicitly as a degree of freedom.  Since the $\eta'_d$ is a new composite state that appears in the low-energy theory, matching is accomplished by identifying symmetry currents which are shared between the dark matter sector and the low-energy chiral theory, specifically the axial current corresponding to U$(1)_A$ chiral rotations.  The dark-quark axial current is $j_A^\mu = \bar{\Psi}_n \gamma^\mu \gamma^5 \Psi_n$, while in the low-energy theory the corresponding axial current is \cite{Scherer:2005ri}
\begin{equation}
j_A^\mu = -f_{\eta'} \partial^\mu \eta_d',
\end{equation}
where $f_{\eta'}$ is the decay constant for the $\eta_d'$, which will scale as $\sqrt{N_D}$ in the large-$N_D$ limit, see \cite{Kaiser:1998ds,Kaiser:2000gs}.  Substituting this for the axial current in \cref{eq:Lint_EW_symmetric}, the resulting low-energy interaction is 
\begin{equation} 
\mathcal{L}_{\eta'} \supset -\frac{c_Z'}{\Lambda^2}  f_{\eta'} \partial_\mu \eta'_d (H^\dagger i D^\mu H + \hc).
\label{eq:Lint_eta_Higgs}
\end{equation}
Integrating by parts and applying equations of motion (see \cref{app:alp_coupling} and \cite{Bauer:2016zfj}), this becomes a direct coupling to Standard Model fermions:
\begin{equation}
\mathcal{L}_{\eta'} \supset  \frac{c_Z'}{\Lambda^2}  f_{\eta'} \eta'_d \left(1 + \frac{h}{v} \right) \sum_f m_f \bar{f} i \gamma_5 f 
\label{eq:Lint_eta_fermions}
\end{equation}
where the index $f$ runs over all Standard Model quarks and leptons.  

Notably, the decay modes of the $\eta'_d$ resulting from this interaction will also be proportional to the mass of the SM fermion $f$, leading to preferential decay of the $\eta'_d$ into the heaviest SM state that is kinematically allowed.  This is a well-known effect in the physics of axion-like particles, as well as in analogous decays within the meson sector of the SM.  We also note that the way in which this coupling is generated leads to suppression by an additional power of the heavy scale $\Lambda$ compared to naive expectations that scalar decays would be mediated by a dimension-5 operator; this can lead to very long lifetimes for the $\eta'_d$ and therefore meaningful phenomenological constraints from BBN, which we will consider below.

There are additional operators in the low-energy effective theory which we have not considered, for example a magnetic moment operator for the light fermion,
\begin{equation} 
\mathcal{L} \supset c_m \frac{\overline{\Psi}_n \sigma^{\mu \nu} \Psi_n F_{\mu \nu}}{\Lambda}
\label{eq:mag_moment_eft}
\end{equation}
where $F_{\mu \nu}$ is the photon field-strength operator.  This operator may be present and lead to significant constraints from direct-detection experiments in the most general version of this model. Below, we estimate that $c_m$ is very small from integrating out the dark equilibration sector that we consider, and is negligible relative to the other couplings that we have identified.  However, this and other operators we have neglected could potentially be important to consider in other realizations of the low-energy dark matter sector.

\subsection{The Dark Equilibration Sector}
\label{sec:equilibration}

The dark equilibration sector serves two purposes: i) it permits the dark sector to be thermalized with the SM at high temperatures, and ii) integrating out the equilibration sector leads to the higher dimensional operators, \cref{eq:Lint_EW_symmetric}, that connect the dark sector with the SM\@.  
As shown in Table~\ref{tab:HSDM-charges}, the particular equilibration sector that we consider consists of two additional Dirac fermions written in terms of left-handed Weyl fermions.
\begin{table}[tp]
\centering 
\renewcommand{\arraystretch}{1.2}
\hspace*{-0.5em}
\begin{tabular}{c|c|c|c|c|c}
              & Field       & $SU(N_D)$             & $(SU(2)_L, Y)$ & $T_3$ & $U(1)_{\rm em}$ \\ \hline  
dark matter   & $n_d$       & ${\bf N}$             & ({\bf 1}, 0)   & 0 & 0  \\
sector        & $n'_d$ & ${\bf \overline{N}}$  & ({\bf 1}, 0) & 0 & 0 \\ \hline  
$\begin{array}{c}
{\rm dark} \\ \vspace{-2mm} \\
{\rm equilibration} 
\end{array}$& $l_d$       & ${\bf N}$             & ({\bf 2}, -$\frac{1}{2}$) &   $\left( \begin{array}{c} +\frac{1}{2} \\ -\frac{1}{2} \end{array} \right)$ & $\left( \begin{array}{c} 0 \\ -1 \end{array} \right)$ \\
{\rm sector}& $l'_d$ & ${\bf \overline{N}}$  & ({\bf 2}, +$\frac{1}{2}$) &  $\left( \begin{array}{c} +\frac{1}{2} \\ -\frac{1}{2} \end{array} \right)$ & $\left( \begin{array}{c} +1 \\ 0 \end{array} \right)$ 
\end{tabular}
\caption{The dark matter and dark equilibration sectors of HSDM model in terms of Weyl (2-component) fermion fields. Note that the heavy fields $l_d, l'_d$ have EM charge-neutral components that can mix with the $n_d$ after electroweak symmetry breaking.  The electric charge is equal to $Q = T_3 + Y$.}
\label{tab:HSDM-charges}
\end{table}
The two Weyl fermions
$l_d$, $\bar{l}_d$ transform as a (fundamental, antifundamental)
under the dark gauge group and also with SM quantum numbers that are
equivalent to the SM lepton doublet with hypercharge $Y = (-1/2, +1/2)$.
From Table \ref{tab:HSDM-charges}, one can easily check that
all gauge anomalies cancel.  

Vector-like masses are permitted for both dark sector field and the fields in the equilibration sector
\begin{equation}
\mathcal{L} \; \supset \;  
-\mnb n_d n'_d + \mlb \epsilon_{ij} l_d^i l'_d{}^j 
+ h.c. \, ,
\label{eq:Vmass}
\end{equation}
where $i,j$ denote SU$(2)_L$ indices which can be raised/lowered by the $\epsilon_{ij}$ tensor as usual, with $\epsilon_{12} = -1$. 
However, what is crucial about the electroweak charges of the equilibration
sector is that they also permit interactions with the Higgs doublet
$H({\bf 2}, +1/2)$ which will lead to additional ``off-diagonal'' mass terms,
\begin{eqnarray}
\mathcal{L} &\supset &
y_{ln}     \epsilon_{ij} l_d^i H^j n'_d
-y_{ln}'  l'_d{}^i H^\star_i n_d
+ h.c. \, .
\label{eq:Hmass}
\end{eqnarray}
In terms of discrete symmetries, we see that charge conjugation $C$ and parity $P$ act as follows:
\begin{eqnarray}
C: & & l_d \leftrightarrow l_d' \; 
\; , \;\; n_d \leftrightarrow n_d' \, , \\
P: & & l_d \leftrightarrow l_d'{}^\dagger \; , \; n_d \leftrightarrow n_d'{}^\dagger,
\end{eqnarray}
as well as the Hermitian conjugates of these relations.
Applying $C$ and $P$ to the Yukawa interactions
involving $l_d$ and $n_d$, we see that
\begin{eqnarray}
C[\epsilon_{ij} l_d^i H^j n'_d] &=& -l'_d{}^i H^\star_i n_d, \\
P[\epsilon_{ij} l_d^i H^j n'_d] &=& \epsilon_{ij} (l'_d{}^{i})^\dagger H^j n_d^\dagger \, .
\end{eqnarray}
The extra minus sign under $C$ comes from the action of charge conjugation on the Higgs field, $H^i \rightarrow \epsilon^{ij} H^\star_j$.  Thus, $C$ invariance requires $y_{ln} = y_{ln}'$, while $P$ invariance requires $y_{ln} = y_{ln}'{}^\dagger$.  In our analysis that follows, we
will take the Yukawa couplings to be real so that $CP$ invariance is preserved.  
However, we require $y_{ln} \neq y_{ln}'$, so that $P$ (and $C$) are not conserved by the Yukawa couplings.  This is required in order to generate a nonzero coefficient for the operator $c_Z'$ in \cref{eq:Lint_EW_symmetric} which, as we will see, allows the $\eta'_d$ meson to decay.  This is consistent with the picture in terms of discrete symmetries here, because the $c_Z'$ operator  manifestly violates parity.

After electroweak symmetry breaking, we can combine the vector-like and electroweak symmetry breaking mass terms from eqs.~(\ref{eq:Vmass}),(\ref{eq:Hmass}), to give the full mass spectrum.  (This was studied in detail in the previous case of stealth dark matter \cite{Appelquist:2015yfa}. \footnote{We note that the present model can be viewed as a limit of stealth dark matter \cite{Appelquist:2015yfa}, with different charge assignments, in which one Dirac fermion is taken to be very light and one Dirac fermion, which would be an $SU(2)_L$ singlet that carries hypercharge -1, is taken infinitely heavy and decoupled.}) There are two Dirac fermions
$\Psi_{n,N}$ with electric charge $Q = 0$
and one Dirac fermions $\Psi_{E}$ with electric charge $Q = -1$.  The neutral fermion masses are
\begin{equation}
m_{n,N}^{\rm full} \, =  \, \frac{\mlb + \mnb}{2} \mp \left[ \left( \frac{\mlb - \mnb}{2} \right)^2 + \frac{y_{ln} y_{ln}' v^2}{2} \right]^{1/2} 
\label{eq:neutralfermion}
\end{equation}
where the superscript ``full'' indicates that this expression is valid for arbitrary vector-like and electroweak symmetry breaking contributions to the masses.  We will be interested in the limit where the equilibration sector fermions are much heavier than the dark sector sector fermion,
\begin{equation}
m_{\rm eq} \; \equiv \; m_{l,0} \gg \big\{ m_{n,0}, \; y_{ln} v, \; y_{ln}' v \big\} \, .
\end{equation}
Expanding in powers of $m_{n,0}/m_{\rm eq}$, we find
\begin{eqnarray}
m_{n}^{\rm full} &\approx& \mnb - \frac{y_{ln} y_{ln}' v^2}{2 m_{\rm eq}} + \label{eq:mn_full} ... \\
m_{N}^{\rm full} &\approx& m_{\rm eq} + \; \ldots \, ,
\end{eqnarray}
and $m_E = m_{\rm eq}$.

Finally, we note that the fermion mixing, 
given by \cref{eq:neutralfermion}, arises 
through tree-level mixing with 
the equilibration sector, Table~\ref{tab:HSDM-charges}, because the equilibration sector 
contains one set of fermions with 
zero electric charge \footnote{One can understand this by recognizing 
that the doublets $l_d$ and $n_d'$ 
have the same electroweak quantum numbers as a left-handed doublet
and right-handed singlet in the SM, 
except that both of these states also transform
under the dark color group.  
Since the equilibration sector is vector-like,
we also have $l'_d$ and $n_d$
that have the opposite quantum numbers
to $l_d$ and $n_d'$ respectively.
The Yukawa interactions
$y_{ln}$ and $y_{ln}'$ mix the neutral 
components of these doublets with the singlets,
and are ultimately responsible for the suppressed
interactions that the light fermion $\Psi_n$ 
has with the SM\@.}.  This is shown in \cref{fig:HHcurrent}.
  Equilibration sectors
without this tree-level mixing would still
allow interactions of the dark sector
with the SM, but it would be suppressed by
one or more loop factors.  These loop factors
would have the effect to further suppress the
interactions of the SM with the dark sector,
leading to much smaller direct detection 
cross sections and much longer lifetimes
for the light mesons such as the $\eta'_d$.

\subsection{Matching onto the EFT}

\begin{figure}
\begin{center}
\includegraphics[width=0.35\textwidth]{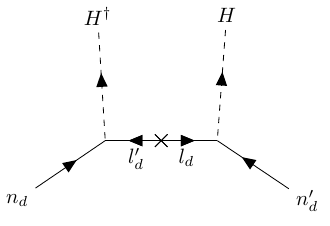}    \end{center}
\caption{Induced Higgs scalar current from integrating out the equilibration sector.  Feynman diagrams in this paper are drawn for two-component Weyl spinors, following \cite{Dreiner:2008tw,Martin:2012us}; there is also a conjugate diagram which is omitted.  The cross on the middle (clashing) propagator indicates an insertion of the vector-like mass $m_{l} \sim \meq$.}
\label{fig:HHcurrent}
\end{figure}

In this section, we consider integrating out the equilibration sector to obtain explicit matching onto the effective couplings $c_s, c_Z, c_Z'$ in terms of the parameters of the UV completion.  The scale suppressing the operators is simply the vector-like mass scale of the equilibration sector:
\begin{eqnarray}
    \Lambda &=& m_{\rm eq} \, ,
\end{eqnarray}
up to corrections of order $y_{ln} v/m_{\rm eq}$.  For example, we can see that the scalar operator $c_s$ in \cref{eq:Lint_EW_symmetric} is generated from the Yukawa couplings $y_{ln}$ and $y_{ln}'$, following the diagram in \cref{fig:HHcurrent}, leading to the matching result
\begin{equation}
c_s = -y_{ln} y_{ln}'.
\end{equation}
This can be easily verified by checking the Higgs contributions to the mass in \cref{eq:mn_full} against \cref{eq:mn_simple}.

\begin{figure}
\begin{center}
\includegraphics[width=0.40\textwidth]{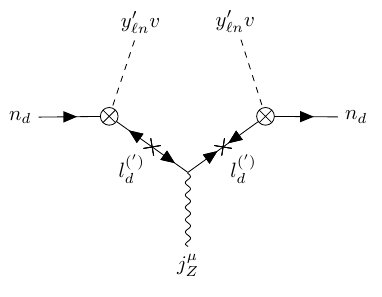}    
\includegraphics[width=0.40\textwidth]{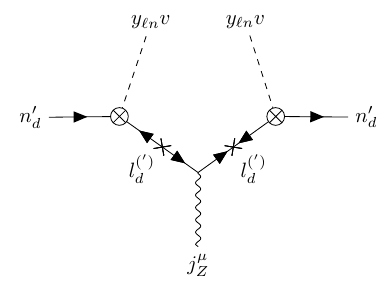}    
\end{center}
\caption{Induced electroweak current from integrating out the equilibration sector, again in two-component notation (see \cref{fig:HHcurrent}.)  Crossed circles indicate insertions of the Higgs vev.  The Yukawa couplings in \cref{eq:Hmass} allow $n_d,n'_d$ to mix with the neutral components of the heavy doublet $l_d, l'_d$.  The two diagrams shown are related by parity, and are equal for a purely vector-like coupling to the $Z$-boson; this makes it manifest that the axial-vector coupling $c_Z'$ requires $y_{ln} \neq y_{ln}'$.  In the limit that the momentum flowing in the diagram is much smaller than $\meq$, the result is dominated by the contribution with mass insertions on both internal propagators, leading to a $1/\meq$ dependence from each, which matches on to \cref{eq:theta_Z_match}.}
\label{fig:Zcurrent}
\end{figure}

To match on to the other effective operators, the simplest way to proceed is in terms of electroweak currents.\footnote{Here we cannot translate directly from \cite{Appelquist:2015yfa}, since the electroweak charge assignments are different in the present case.}  For the singlet field $\Psi_n$, we find the result
\begin{equation}
j_Z^\mu \, = \, j_3^\mu \, = \, \overline{\Psi}_n \gamma^\mu \left( \sin^2 \theta_1 P_L + \sin^2 \theta_2 P_R \right) \Psi_n, \\
\end{equation}
where $\theta_{1,2}$ are mixing angles that arise from integrating out the equilibration sector.  This can be done in general (e.g., see \cite{Appelquist:2015yfa}), but for our purposes, it is most easily understood from Fig.~\ref{fig:Zcurrent}, 
where we use the mass insertion approximation and take the heavy equilibration sector fermions to have mass $\Lambda = m_{\rm eq}$, we can then read off
\begin{equation} \label{eq:theta_Z_match}
\theta_1 \, \approx \, \frac{y_{ln}' v}{\sqrt{2} m_{\rm eq}} \, , \quad \theta_2 \, \approx \, \frac{y_{ln} v}{\sqrt{2} m_{\rm eq}}.
\end{equation}
Matching onto the low-energy theory, from \cref{eq:nZ_current} the corresponding Z current is 
\begin{equation}
j_Z^\mu \, = \, \frac{\sqrt{g^2 + g'^2} v^2}{2m_{\rm eq}^2} \left( c_Z \bar{\Psi}_n \gamma^\mu \Psi_n + c_Z' \bar{\Psi}_n \gamma^\mu \gamma^5 \Psi_n \right)
\end{equation}
and thus we identify
\begin{eqnarray}
c_Z &=& \frac{(y_{ln})^2 + (y_{ln}')^2}{2\sqrt{g^2 + g'^2}}, \\
c_Z' &=& \frac{(y_{ln})^2 - (y_{ln}')^2}{2\sqrt{g^2 + g'^2}}.
\end{eqnarray}
Defining 
\begin{eqnarray}
y_{ln} &=& y (1 + \epsilon) 
\label{eq:yln} \\
y_{ln}' &=& y(1 - \epsilon) \, ,
\label{eq:ylnp} 
\end{eqnarray} 
then we can rewrite these as
\begin{eqnarray}
c_Z &=& \frac{y^2 (1 + \epsilon^2)}{\sqrt{g^2 + g'^2}}, \\
c_Z' &=& 2\epsilon^2 c_Z.
\end{eqnarray}
It is helpful to rewrite once more in terms of the simplified mixing angle
\begin{equation} 
\theta \, \equiv \, \frac{yv}{\sqrt{2} m_{\rm eq}},
\label{eq:def_theta}
\end{equation}
which for the three effective couplings leads to:
\begin{eqnarray}
\frac{c_Z}{\Lambda^2} &=& 
\frac{c_Z}{m_{\rm eq}^2} \; = \;  
\theta^2 \frac{2(1+\epsilon^2)}{\sqrt{g^2 + g'^2}v^2} \;=\; \frac{\theta^2 (1 + \epsilon^2)}{2 M_Z^2}, \label{eq:cz_theta} \\
\frac{c_Z'}{\Lambda^2} &=& 
\frac{c_Z'}{m_{\rm eq}^2} \;=\; \frac{\epsilon^2 \theta^2}{M_Z^2}, 
\label{eq:czprime_theta} \\
\frac{c_s}{\Lambda} &=& 
\frac{c_s}{m_{\rm eq}} \;=\; -\sqrt{2} 
\frac{\theta}{v} y \; = \; -2 \frac{\theta^2}{v} \frac{m_{\rm eq}}{v}. \label{eq:cs_theta}
\end{eqnarray}
%
%

Finally, let us return to briefly consider the magnetic moment operator $c_m$, \cref{eq:mag_moment_eft}.  This will be induced by mixing of the light $\psi_n$ fermion into the heavy neutral state, which can then emit a $W$ boson to couple to the photon.  The resulting magnetic moment will be proportional to $\alpha \theta^2$, similar to the $Z$ coupling.  Moreover, the dark baryon formed from $\psi_n$ will be expected to have a similar overall magnetic moment.  However, the resulting cross section for direct detection of dark baryons via magnetic moment is proportional to $\alpha^4 \theta^4$ \cite{LatticeStrongDynamicsLSD:2013elk}, which is sufficiently suppressed relative to $Z$ exchange (see \cref{sec:direct} below) that it can be safely neglected for our purposes.

\section{\label{sec:model} Model parameters,  mass scales, and constraints}

Before discussing the phenomenology in detail, let us briefly review the free parameters of the Hyper Stealth Dark Matter model.  For the low-energy theory, there is the number of dark colors $N_D$, the dark fermion mass $m_n$, and the dark confinement scale $\Lambda_d$.  Additional parameters which are introduced by way of the equilibration sector (the UV completion) are the scale $m_{\rm eq}$ of the heavy fermions, the mass-mixing parameter $\theta$, and the parity-violating parameter $\epsilon$.  These six parameters fully specify the theory under the assumptions we have made.

\subsection{Confined Low-Energy Description with $N_D$ Dark Colors}
\label{sec:confinedND}


In terms of phenomenology, a number of  other quantities are of direct
significance, including the dark matter mass $M_{B_d}$, the dark meson
mass $M_{\eta'_d}$, the decay constant $f_{\eta'_d}$, and other light
meson masses such as the vector meson mass $M_{\rho_d}$, the lightest glueball and its associated decay constant.  In principle, these are predictions of the strong dynamics that may be determined based on lattice calculations at a given $N_D$ and $m_n / \Lambda_d$ (with the third parameter $\Lambda_d$ simply fixing the overall energy scale.)  In the absence of specific lattice results, we need rough estimates in order to proceed.  We adapt the results of \cite{DeGrand:2019vbx}, based on lattice calculations for SU(3) at a variety of quark masses, and apply large-$N$ scaling relations \cite{Lucini:2012gg, Giacosa:2024scx}.  Based on the predictions at intermediate and heavy quark masses (specifically, we impose $M_{\eta'_d}^2 / M_{\rho_d}^2 \gtrsim 0.1$ to avoid going to too-light fermion masses where the one flavor theory should diverge greatly from the behavior of QCD since the $\eta'_d$ is not a pseudo-Goldstone boson), we adopt the following relations:
\begin{eqnarray}
  \frac{M_{B_d}}{M_{\eta'_d}} &=& \frac{N_D}{2}, \\
  \label{eq:quark_ratio}
\frac{f_{\eta'_d}}{M_{\eta'_d}} &=& \frac{1}{2} \sqrt{\frac{N_D}{3}},
\end{eqnarray}
which will be used to obtain numerical results below.  We emphasize that these are phenomenological estimates from lattice results, not strictly large-$N_D$ predictions.  This value for the ratio $f_{\eta'_d}/M_{\eta'_d}$ is fairly robust over the large range of fermion masses considered.  On the other hand, the ratio of the masses of $B_d$ to $\eta'_d$ has a slightly larger variation over the range
\begin{equation}
\frac{N_D}{3} \, \lesssim \, \frac{M_{B_d}}{M_{\eta'_d}} \, \lesssim \, \frac{2 N_D}{3} \, .
\label{eq:mBovermetap}
\end{equation}
For this ratio of masses, we will adopt the fiducial value of $N_D/2$ as the central value.  In cases where the dependence on $M_{\eta'_d}$ is significant, we will estimate the uncertainty by varying over the larger range given by \cref{eq:mBovermetap}.

The masses of other intermediate mesons will also be useful to estimate; in particular, the $\sigma$ meson, with $0^{++}$ quantum numbers, will be relevant for one of the bounds to be discussed below. An early lattice study of this state in QCD \cite{Kunihiro:2003yj} found $M_\sigma \gtrsim M_\rho$ at relatively heavy quark masses.  For simplicity, we will take $M_{\sigma_d} \approx M_{\rho_d}$ and then use the results for the latter from \cite{DeGrand:2019vbx}, finding that the simple large-$N_D$ result works reasonably well over the full mass range:
\begin{equation} \label{eq:quark_mass_range}
\frac{M_{B_d}}{M_{\rho_d}} \, \approx \,  \frac{M_{B_d}}{M_{\sigma_d}} \, \approx \, \frac{N_D}{2}. 
\end{equation}

This parametrization allows us to fix the other quantities in terms of the dark matter mass $\mdm$, which we use to set the overall mass scale.  However, we do not have direct access to the light fermion mass $m_n$ in this approach. From \cite{DeGrand:2019vbx}, the lower end of the mass range that we consider ($M_{\eta'_d}^2 / M_{\rho_d}^2 \sim 0.1$) corresponds (in QCD) to a quark mass of approximately $20$~MeV, with a nucleon mass of about $1070$~MeV\@.  In the heavy-quark limit, the ratio $\mdm / m_n$ should approach $N_D$.  Assuming the other limit also scales proportional to $N_D$, we consider the range
\begin{equation}
N_D \, \lesssim \, \frac{\mdm}{m_n} \, \lesssim \, 18 N_D.
\end{equation}
Further details on how this and the other phenomenological estimates above are obtained in a data-driven way can be found in the ancillary material.\footnote{Python code used to study hadron mass ranges and to create all of the plots in this paper can be found at \url{https://github.com/etneil/hsdm_code}.}

Finally, we will be interested in the mass of the $0^{++}$ glueball state.  Again based on the summary in \cite{DeGrand:2019vbx}, little is known about glueball masses in theories with dynamical fermions as a function of the fermion mass.  In the light-quark limit, we expect very roughly that $M_{0^{++}} \gtrsim M_{B_d}$, similar to the case in QCD, since both masses will be fixed by the confinement scale.  On the other hand, in the heavy fermion limit $m_n \gg \Lambda_d$ we expect that $M_{0^{++}} \ll M_{B_d}, M_{\eta'_d}$; by increasing the ratio $m_n / \Lambda_d$, the glueball can be made arbitrarily light compared to the other hadronic states.  In the absence of more quantitative information, we will simply consider the two cases $M_{0^{++}} \approx M_{B_d}$ (``light-fermion limit'': $m_n \ll \Lambda_d$) and $M_{0^{++}} \approx M_{\eta'_d}/2$ (``heavy-fermion limit'': $m_n \gg \Lambda_d$) \footnote{Taking $M_{0^{++}}$ much lighter than $M_{\eta'_d}$ is possible, but will tend to put the model into a heavily constrained part of parameter space, as will be described below.}.

\subsection{Precision Electroweak Constraints} 
\label{subsec:pew} 

As we have seen, the presence of Higgs interactions 
between the dark sector and the 
equilibration sector causes mass mixing
between the electroweak singlets
of the dark sector and the electroweak
doublets of the equilibration sector.
This means that the dark sector confinement
is not completely innocuous with respect
to electroweak symmetry breaking.
Namely, we expect the additional states will lead 
to a small amount of electroweak
symmetry breaking following the condensation of the dark sector fermions.  In more general cases, one could attempt to estimate contributions to electroweak precision in terms of the EFT couplings such as $c_Z$ and $c_S$, but here we consider the equilibration sector contributions directly instead.  

The electroweak symmetry breaking 
inherent in the combined dark matter
and equilibration sectors 
give small but nonzero contributions
to the electroweak precision parameters
$S,T,U$ \cite{Peskin:1991sw}.  These contributions
arise because the Yukawa interactions,
eqs.~(\ref{eq:Hmass}), cause a small
mixing between the light dark sector
states with the equilibrium sector states.
A precise calculation would require taking
into account the nonperturbative effects
of the dark sector confinement,
but this is beyond the scope of the paper.
We can, however, obtain a perturbative 
estimate of the $S$ and $T$ parameters 
by including 
the effects of the additional states 
scaled by the number of dark colors
using the results in \cite{Lavoura:1992np}.
Even the perturbative contribution is 
somewhat opaque when expressed in terms
of the full set of parameters of the
theory.  If we work in the approximations
that all heavy states in the equilibration sector have mass $m_{\rm eq} \gg M_Z$, the one light flavor has mass $m_n \ll M_Z$, 
and take $\epsilon$ small, we obtain
\begin{eqnarray}
    S &\simeq& \frac{\xi_S N_D}{\pi} \theta^2 \; \simeq \; 0.3 \, \xi_S N_D \theta^2 \\
    T &\simeq& \frac{N_D}{8 \pi \cos^2 \theta_W \sin^2\theta_W} \frac{m_{\rm eq}^2}{M_Z^2} \theta^4 \; \simeq \; 0.8 \, y^2 N_D  \theta^2 
\end{eqnarray}
where $\xi_S$ is a complicated kinematic function of the light and heavy masses.  For the ranges of parameters considered in this paper, we find $0.5 \lesssim \xi_S \lesssim 2$.  Given perturbative Yukawa couplings $y \lesssim 1$, and the small $\theta \ll 1$ considered 
throughout this paper, we find the contributions to $S,T$ to be well within the limits set by precision electroweak data.  The nearest approach to a constraint
would arise if $y \simeq 1$,
$\theta = 0.1$ (the largest we consider in
this paper), and then $T \lesssim 0.1$
implies $N_D \lesssim 12$.  Since the constraint
on the number of dark colors scales as 
$\theta^{-2}$, it rapidly 
disappears as $\theta$ is taken well below $0.1$.

\subsection{Constraints from Higgs Mixing} 
\label{subsec:higgsmixing} 

Additional effects to precision data arise from dark sector states mixing with the visible sector states.  
The largest mixing arises from 
the dimension-5 $\overline{\Psi}_n \Psi_n H^\dagger H$ interaction in \cref{eq:Lint_EW_symmetric}.
Once the dark sector confines, 
the $\sigma$ state appears as $\langle 0|\bar{\Psi}_n\Psi_n|\sigma\rangle \equiv \mathbf{F_\sigma}$ in terms of the non-perturbative decay constant $\mathbf{F_\sigma}$ with mass dimension two.
This leads to  sigma-Higgs mixing
\begin{eqnarray}    
\frac{c_s v \, \mathbf{F_\sigma}}{\Lambda} \sigma \, h  & \simeq &  \sqrt{2} \, \mathbf{F_\sigma} \, y \, \theta \, \sigma \, h \, ,
\end{eqnarray}
that can be characterized by a mixing angle
\begin{eqnarray}
    \tan 2 \gamma &=& \frac{2 \sqrt{2} \, \mathbf{F_\sigma} \, y \, \theta}{m_\sigma^2 - m_h^2} \, .
\end{eqnarray}
When $m_\sigma \ll m_h$, Ref.~\cite{Winkler:2018qyg}
found constraints from scalar emission and decay
back into SM states can reach $\sin \gamma \sim 10^{-4}$
for $m_\sigma \lesssim 5$~GeV\@. 
Assuming $\mathbf{F_\sigma} \sim m_\sigma^2$, this leads
to the constraint 
$y \, \theta \lesssim 0.05$ for $m_\sigma \sim 5$~GeV, 
and correspondingly weaker constraints for smaller
$m_\sigma$.
For larger sigma masses, there are LHC bounds from 
precision Higgs couplings as well as direct searches.
Using the results from \cite{Buttazzo:2018qqp},
we estimate the 
current sensitivity $\sin \gamma \sim 0.1$-$0.3$.
In the regime $m_\sigma > m_h$, these bounds 
restrict $y \theta \lesssim 0.1$. 
We therefore find that when $\theta \lesssim 0.1$,
constraints from sigma-Higgs mixing are 
satisfied.  Future colliders have the ability
to probe considerably smaller mixings \cite{Buttazzo:2018qqp}
providing a fascinating opportunity to 
investigate the scalar mesons that appear in
this model.

\subsection{Fine-tuning Constraints from Electroweak Symmetry Breaking}
\label{subsec:ewsbfinetuning}

There are two effects to consider
in the dark sector EFT\@.  
If dark sector confinement occurs
prior to electroweak symmetry breaking,
i.e., $\Lambda_d > v$, the dimension-5 operator in \cref{eq:Lint_EW_symmetric}
will lead to a contribution to the
Higgs doublet $(\mbox{mass})^2$,
\begin{eqnarray}
    m_{H,{\rm tot}}^2 &\simeq& m_{H,{\rm tree}}^2 + c_s \frac{\Lambda_d^3}{\Lambda} \nonumber \\
          &\simeq& m_{H,{\rm tree}}^2 + y_{ln} y'_{ln} \frac{\Lambda_d^3}{m_{\rm eq}}
\end{eqnarray}
This is simply an additive shift to the effective potential for the Higgs doublet.
While the Higgs $(\mbox{mass})^2$ 
is already infamously quadratically sensitive
to new physics, we will simply assume
there is no excessive tuning
between these two contributions,
and thus require
\begin{eqnarray}
   y_{ln} y'_{ln} \frac{\Lambda_d^3}{m_{\rm eq}} &\lesssim& v^2 \qquad [\Lambda_d > v] \, ,
\end{eqnarray}
or equivalently
\begin{eqnarray}
   \theta &\lesssim& \frac{v^2}{(\Lambda_d^3 m_{\rm eq})^{1/2}} \qquad [\Lambda_d > v] \, .
\end{eqnarray}
For numerical study of the resulting bound, we use $\Lambda_d \sim M_{0^{++}_d}$

In the regime where $\Lambda_d < v$, electroweak symmetry breaking in the Higgs sector contributes to the vector-like mass of the dark sector vectorlike fermion mass.
This causes the shift 
\begin{equation}
m_n \approx \mnb  - \frac{y_{ln} y'_{ln} v^2}{2\meq} = \mnb - \theta^2 \meq \qquad [\Lambda_d < v] \, .
\end{equation}
From collider searches for composite electroweak particles,  e.g.~\cite{Kribs:2018ilo,Butterworth:2021jto}, we require $\meq \gtrsim 3$ TeV.  For the contributions to the mass $m_n$, which we take to be much smaller than the TeV scale, if $\theta$ is too large then some fine-tuning will be introduced.  For example, suppose that we want to obtain $m_n \sim 1$ GeV with $\meq \sim 3$ TeV\@.  If $\theta = 0.1$, then $\theta^2 \meq \sim 30$ GeV, and fine-tuning of $\mnb$ to 1 part in 30 is required.

In order to avoid the fine-tuning between vectorlike mass and the electroweak symmetry breaking contributions, 
we require
\begin{equation}
\theta^2 \meq \, \lesssim \, m_n \, .
\end{equation}
This ensures we are not relying on any significant cancellation between bare and Higgs-induced mass terms in order to obtain the physical $n$ mass.  This leads to a disfavored region shown that we will show below in our numerical plots in the light fermion mass limit $M_N \sim 18 N_D m_n$ (see discussion around \cref{eq:quark_mass_range}).

\section{\label{sec:direct} Direct detection}

%
\begin{figure*}
\begin{center}
\includegraphics[width=0.49\textwidth]{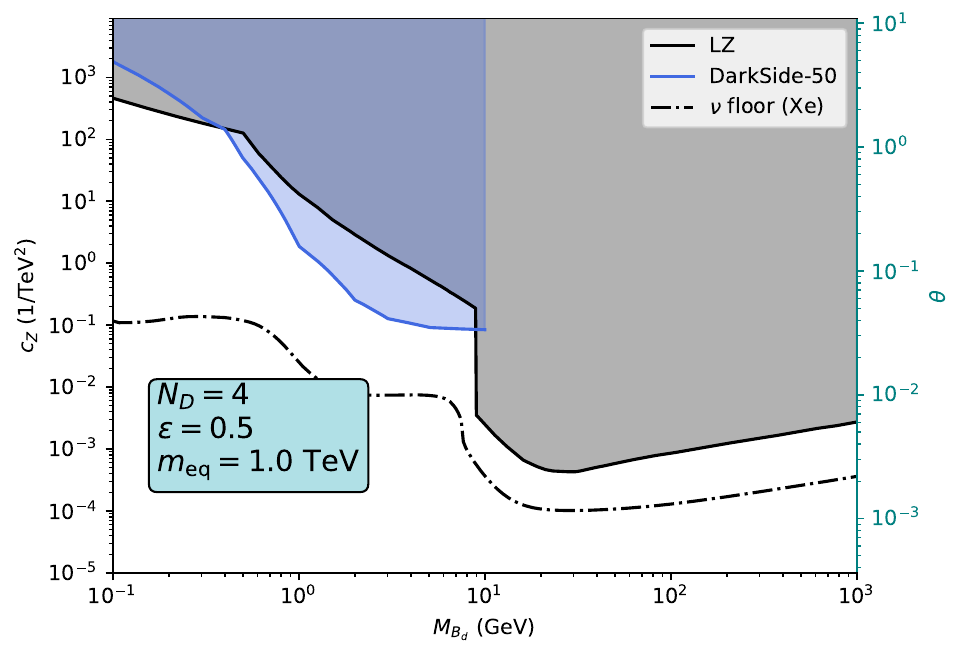}
\includegraphics[width=0.49\textwidth]{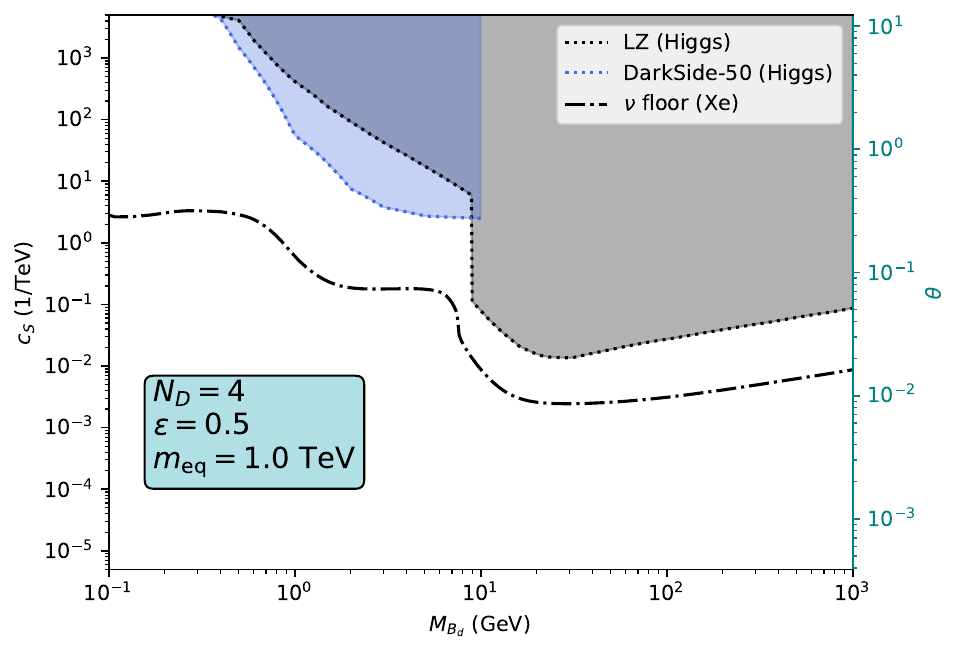}
\end{center}
\caption{Left: Bounds from direct-detection experiments LZ \cite{LZ:2022lsv,LZ:2023poo} and DarkSide-50 \cite{DarkSide:2022dhx}, as discussed in \cref{sec:direct}.  The left axis shows the bound on the generic EFT parameter $c_Z$, in $\textrm{TeV}^{-2}$; the right axis shows the equivalent bound on $\theta$ with our dark equilibration sector UV completion.  The dashed line shows projected bounds at the neutrino floor for xenon from \cite{OHare:2021utq}. Right: same as left figure but for the Higgs EFT parameter $c_s$. Note the y-axis scale in terms of $\theta$ is held roughly fixed, showing that the $c_S$ bound is less restrictive than the $c_Z$ bound for our UV completion.}
\label{fig:DD}
\end{figure*}

We now turn to consider the elastic scattering cross section of the HSDM dark baryon with the SM\@. 
Following the derivation of
\cite{Servant:2002hb}, the Z-exchange cross section of the dark baryon $B_d$ with an atomic nucleus with atomic number $Z$ and mass number $A$ is:
\begin{eqnarray}
  \sigma_Z(B_d) &=&
  {\mu^2 G_F^2 \over 2 \pi} [ (1-4 \sin^2{\theta_W} ) Z - (A-Z)]^2 \nonumber \\
  &\times&  |\langle B_d| j_Z |B_d\rangle|^2 
  \label{eq: cross_section_Z_2}
\end{eqnarray}
where $\mu$ is the reduced mass of $\mdm$ and the target nucleus,
$G_F$ is the Fermi constant, and $\theta_W$ is the Weinberg
angle.  Substituting in \cref{eq:baryon_ME_Z,eq:cz_theta}, this becomes
\begin{eqnarray}
  \sigma_Z(B_d) &=&
  {\mu^2 G_F^2 \over 2 \pi} [ (1-4 \sin^2{\theta_W} ) Z - (A-Z)]^2 \nonumber \\
  &\times&  N_D^2 \theta^4 (1 + \epsilon^2)^2 \frac{v^2}{4M_Z^2} \, .
\end{eqnarray}
In the formula above, we are neglecting the spin $J$ of the dark baryon.  As discussed in \cref{sec:hsdm}, while a quark-model argument would predict $J = N_D/2$ for a one-flavor theory such as this, ultimately lattice calculations should be done to verify the spin of the ground-state baryon $B_d$.  In either case, the contribution due to higher spin is expected to be subdominant in non-relativistic scattering, suppressed relative to the spin-independent scattering we consider by $(v_{\rm rel}/c)^2$, where $v_{\rm rel} \sim 10^{-3} c$; see e.g.~Ref.~\cite{Criado:2020jkp} for a calculation for Higgs exchange with arbitrary spin which shows this suppression explicitly.  We therefore neglect the spin-dependent terms for direct detection in this work, although it is possible that interesting and distinctive signatures could arise from spin-dependent scattering operators with higher spin \cite{Gondolo:2021fqo}.

For comparison to direct-detection experimental bounds, we convert to the per-nucleon cross section \cite{Lisanti:2016jxe},
\begin{equation}
\sigma_{Z,a} \, = \, \sigma_Z \frac{\mu_a^2}{A^2 \mu^2} \, ,
\end{equation}
where $\mu_a$ is the reduced mass for the dark baryon with a single nucleon rather than with the nucleus.

We will obtain our bounds using results from the LZ experiment, for which the target nucleus is xenon ($Z=54$, various stable isotopes with $A \approx 130$.)  For $\mdm \geq 9$ GeV, we use the results of \cite{LZ:2022lsv}; in the range $0.5$ GeV $\leq \mdm < 9$ GeV, we adopt the limits placed in \cite{LZ:2023poo} using low-energy electron recoils, interpreted as bounds on WIMP-nucleus scattering via the Migdal effect.  We also show results from the DarkSide-50 experiment \cite{DarkSide:2022dhx}, which also uses electron recoils and the Migdal effect with an argon target ($Z = 18$, $A = 40$).  Finally, we show projected ultimate limits for direct detection at the neutrino floor, from \cite{OHare:2021utq}.

As a simple check on our numerical results, using the properties of xenon and taking $(1 - 4\sin^2 \theta_W) \approx 0$, we find the order of magnitude estimate
\begin{equation}
\sigma_{Z,a} \, \approx \, 10^{-37} \left(\frac{\mu_a}{1\ {\rm GeV}}\right)^2 \theta^4 ~{\rm cm}^2 \, . 
\end{equation}

The dark baryon can also interact through Higgs exchange; from \cite{LatticeStrongDynamicsLSD:2014osp}, the corresponding per-nucleon cross section is
\begin{eqnarray}
\sigma_{H,a}(B_d) &=& \frac{\mu_a^2}{\pi A^2} (Z {\mathcal M}_p + (A-Z) {\mathcal M}_n)^2 \, ,
\label{eq:higgs_cross_section_1}  \\
 {\mathcal M}_a &=& \frac{g_a g_{B_d,h}}{m_H^2} \, ,
\label{Higgs_amplitude_1}
\end{eqnarray}
where $a$ labels either a proton ($p$)  or a neutron ($n$).  The Higgs-nucleon coupling is
\begin{equation}
g_a \, = \, \frac{m_a}{v} \left[ \sum_{q = \{u,d,s\}} f_q^{(a)} + \frac{6}{27} \left( 1 - \sum_{q = \{u,d,s\}} f_q^{(a)} \right)\right] ,
\end{equation}
and the Higgs-dark baryon coupling, computed in \cref{app:trace_anomaly}, is
\begin{equation}
g_{B_d,h} \, = \, \frac{\mdm}{v} \theta^2 \left[ f_n^{(B_d)} \frac{\meq}{m_n} + \frac{4}{11N_D - 2} (1 - f_n^{(B_d)}) \right].
\end{equation}
Using $\meq \sim 5$ TeV, taking the Standard Model nucleon-Higgs couplings from \cite{Alexandrou:2014sha,Freeman:2012ry,FlavourLatticeAveragingGroupFLAG:2021npn}, and neglecting the heavy-quark contribution in $g_{B_d,h}$ which is suppressed by $m_n/\meq$, for xenon we find the order of magnitude estimate
\begin{equation}
\sigma_{H,a} \, \approx \, 5 \times 10^{-39} \left(\frac{\mu_a}{1\ {\rm GeV}}\right)^2 \theta^4 \left[f_n^{(B_d)}\right]^2 \  {\rm cm}^2 \, .
\end{equation}
This is always sub-leading compared to the $Z$ exchange cross-section, so we neglect it in our exclusion plots.

It should be noted that in the estimate above, we assume the ratio $\mdm / m_n \approx N_D$ is held fixed.  If this ratio is enhanced, i.e.~if the theory is pushed towards the regime where $m_n$ is much lighter than the dark confinement scale, then the Higgs exchange cross-section can also be enhanced and may yield the dominant direct-detection bound.  However, we can obtain another estimate by adopting the constraint $\theta^2 \meq \lesssim m_n$, which is a condition to avoid fine-tuning discussed in \cref{subsec:ewsbfinetuning} below.  This implies that (neglecting the heavy-quark contribution again)
\begin{equation}
g_{B_d,h} \, \lesssim \, \frac{\mdm}{v} f_n^{(B_d)},
\end{equation}
resulting in the condition
\begin{equation}
\sigma_{H,a} \, \lesssim \, 1 \times 10^{-47} \left(\frac{\mdm}{1\ {\rm GeV}}\right)^2 \left(\frac{\mu_a}{1\ {\rm GeV}}\right)^2 [f_n^{(B_d)}]^2 \  {\rm cm}^2 \, .
\end{equation}
This will be sub-leading compared to the $Z$ exchange cross-section as long as
\begin{equation}
\sigma_{H,a} \, \lesssim \, \sigma_{Z,a} \; \Rightarrow \; \mdm \, \lesssim \, \frac{\theta^2}{f_n^{(B_d)}} \times (100\ {\rm TeV}) \, .
\end{equation}
For $\mdm \lesssim 10$ GeV and $\theta$ which saturates the fine-tuning restriction, this condition is always met, assuming $f_n^{(B_d)} < 1$.  In other words, for lighter $\mdm$ Higgs exchange is always subleading outside of the fine-tuned region of parameter space and may be neglected, regardless of the ratio $m_n / \mdm$.  For heavier $\mdm$, Higgs exchange may become relevant for very light $m_n \ll \mdm$, but we will neglect it for the bounds shown here.
We show bounds from direct detection 
experiments in \cref{fig:DD}.

\section{\label{sec:bbn} Meson decay and BBN}

\begin{figure*}
\begin{center}
\includegraphics[width=0.49\textwidth]{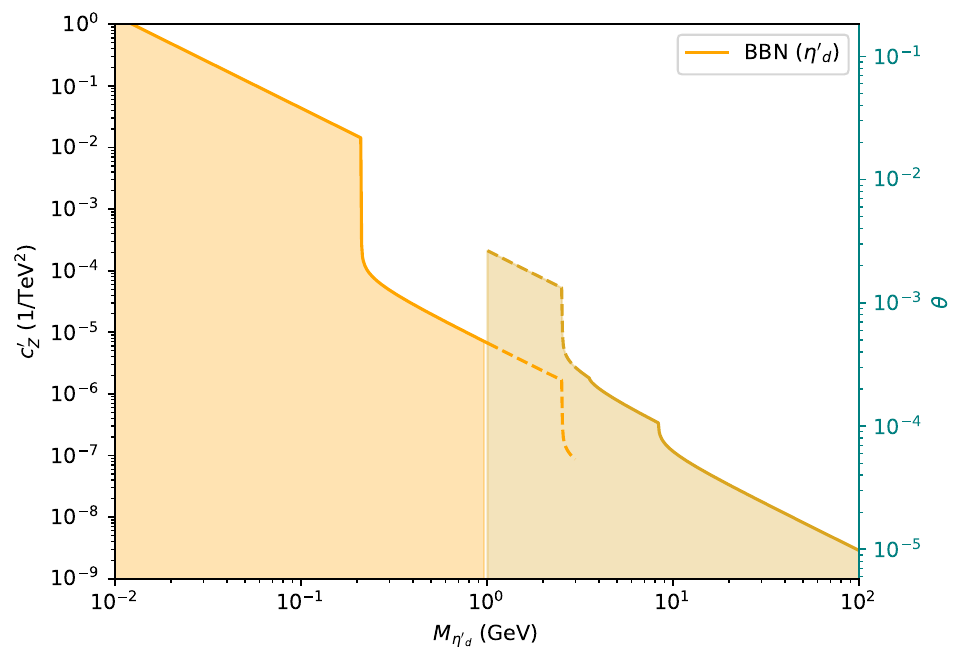}
\end{center}
\caption{Bounds on the $\eta'_d$ meson lifetime from big-bang nucleosynthesis, as discussed in \cref{subsec:eta_decay}.  The left axis shows bounds on the EFT coupling $c_Z'$, in units of TeV${}^{-2}$, while the right axis shows the equivalent bound on $\theta$ with our dark equilibration sector UV completion.  The two separate curves shown correspond to dominant hadronic branching of the $\eta'_d$ ($M_{\eta'_d} > 3$ GeV), and dominant branching to leptons ($M_{\eta'_d} < 1$ GeV).  Both curves are extrapolated into the region between $1-3$ GeV and shown as dashed lines; as discussed in the text, the BBN bound is uncertain here due to hadronic decays near the QCD scale.  Note that there is a region in $M_{\eta'_d}$ from 1 GeV down to $2m_\mu$ where decays to pairs of muons are dominant, and the BBN bound is significantly weakened compared to either lighter or heavier $\eta'_d$ masses.}
\label{fig:BBN_eta}
\end{figure*}

The primary constraint on meson decay is from Big Bang nucleosynthesis \cite{Jedamzik:2006xz,PhysRevD.97.023502}.  Regardless of the details of dark matter relic abundance, dark mesons are expected to be produced abundantly in the early universe, and if they are sufficiently long-lived, their decays can cause observable changes to Big Bang nucleosynthesis (BBN). In principle, a detailed calculation of the relic abundance of $\eta'_d$ and the other mesons is required to study this effect, but this is beyond the scope of this work.  Instead, we will adopt the conservative bound that the decay lifetimes of our dark mesons are shorter than the relevant timescales for BBN.

Following \cite{Jedamzik:2006xz,PhysRevD.97.023502}, for a given dark meson $\phi_d$, in regions of parameter space where the hadronic branching $B_{\phi_d}^h \approx 1$ we adopt the bound $\tau_{\phi_d} \lesssim 0.1$ s.  For all dark mesons, we take this limit to apply for $M_{\phi_d} \geq 3$ GeV, above the $c\bar{c}$ threshold.  Where the hadronic branching drops to zero, typically for much lighter meson masses, we apply the slightly weaker constraint $\tau_{\phi_d} \lesssim 10^2$ s.  This limit is taken to apply for $M_{\phi_d} < 1$ GeV for pseudoscalars, and $M_{\phi_d} < 0.5$ GeV for scalars, since only the latter states may be able to decay into pairs of ordinary pions.  In the region between these two limits, our bounds are somewhat uncertain since we do not have detailed estimates of direct hadronic decays near the QCD scale.  

The meson sector formed from the single light flavor in the dark matter sector 
consists of the $\eta'_d$ meson, as well as additional
(quark-like) mesons formed from this light flavor 
as well as mesons formed from pure glue, otherwise known as dark glueballs.  
The mesons of these sectors do, in general,
mix with one another, complicating
a detailed analysis of their properties.
Nevertheless we expect that the relic abundance of mesons will consist dominantly of just the lightest states; this was explicitly shown for
dark glueballs in Ref.~\cite{Forestell:2016qhc,Forestell:2017wov}.
In this section, in addition to the $\eta'_d$, we consider two additional potentially light meson states:  
the $\sigma$ meson with $J^{PC} = 0^{++}$ formed from 
$\overline{\psi}_n\psi_n$, and the lightest $0^{++}$ glueball meson.

\subsection{$\eta'_d$ decay} \label{subsec:eta_decay}

The $\eta'_d$ interaction Lagrangian \cref{eq:Lint_eta_fermions} is in the standard form for interaction of an axion-like particle (ALP) with the Standard Model, as described in e.g.~\cite{Bauer:2017ris}.  Following the notation of the reference, we identify
\begin{equation}
c_{ff}^{(a)} \; = \; c_Z' \frac{f_{\eta'}}{\meq}.
\end{equation}
We can thus adopt the available results for decay widths of axion-like particles.  For decay into leptons or quarks, we have \cite{Bauer:2017ris}
\begin{eqnarray}
\Gamma(\eta'_d \rightarrow f \bar{f}) &=& N_C^{f} \frac{M_{\eta'} m_f^2}{8\pi \meq^2} \left|c_Z'\right|^2 \frac{f_{\eta'}^2}{\meq^2} \sqrt{1 - \frac{4m_{f}^2}{M_{\eta'}^2}} \\
&=& \frac{N_C^f}{8\pi} M_{\eta'} \theta^4 \epsilon^4 \frac{m_f^2 f_{\eta'}^2}{M_Z^4} \sqrt{1 - \frac{4m_{f}^2}{M_{\eta'}^2}} \, ,
\end{eqnarray}
using \cref{eq:czprime_theta} to substitute for $c_Z'$.  $N_C^f$ is a color factor for the Standard Model fermions, equal to 3 for quarks and 1 for leptons. 
We show the bounds on the parameter using from $\eta'$ decay in \cref{fig:BBN_eta}. 

\begin{figure*}
\begin{center}
\includegraphics[width=0.49\textwidth]{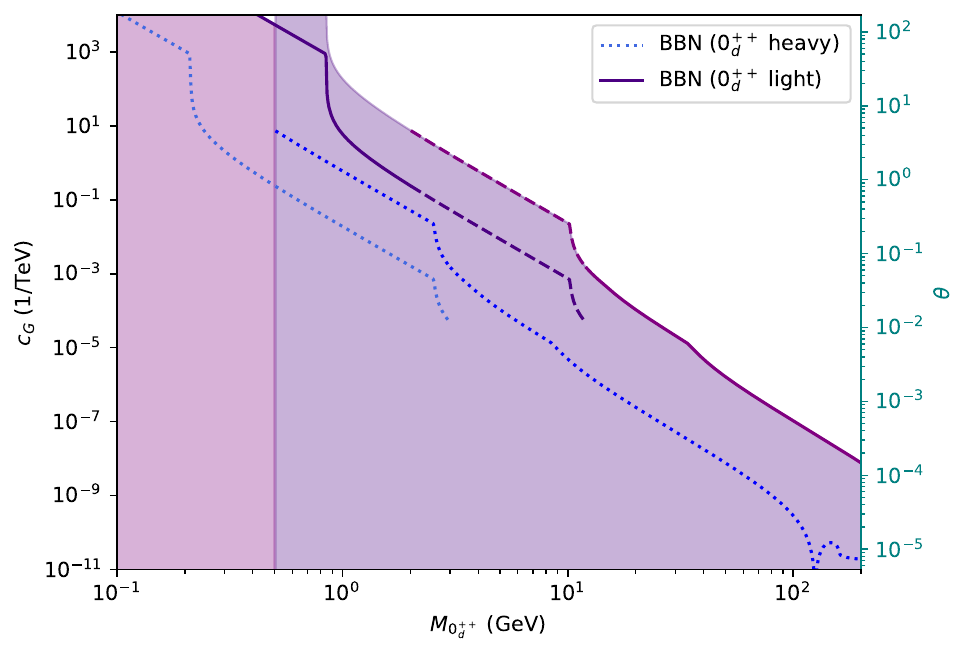}
\includegraphics[width=0.49\textwidth]{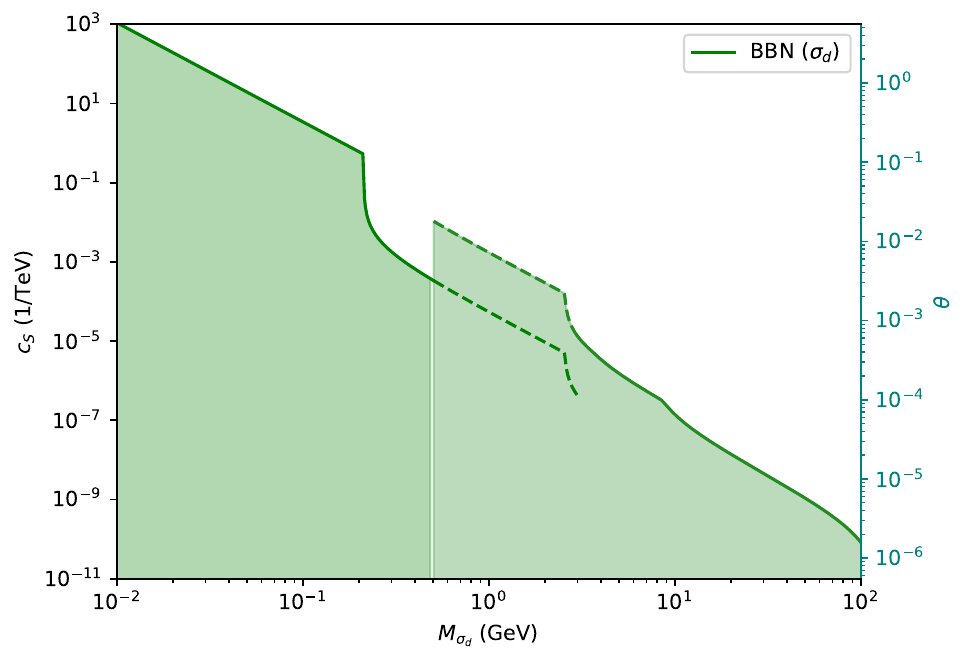}
\end{center}
\caption{Bounds on the scalar meson states $\sigma_d$ and $0^{++}_d$, as described in \cref{fig:BBN_eta}.  For the $0^{++}_d$ dark glueball state,  two sets of bounds are shown: in the heavy-quark limit $m_n \gg \Lambda_d$, the $0^{++}_d$ state is relatively light, giving the ``$0^{++}_d$ light'' case (purple).  In the light-quark limit $m_n \ll \Lambda_d$, the $0^{++}_d$ is comparable to other meson masses, giving the ``$0^{++}_d$ heavy'' case (blue).  For further discussion of these model parameters and mass scales, see \cref{sec:model}.}
\label{fig:BBN_scalar}
\end{figure*}

The above formula is a reasonable description for decays to heavy quarks which, due to the proportionality of this decay mode to the final-state mass $m_f$, will dominate as long as they are kinematically allowed.  For sufficiently light $\eta'_d$, as with more general axion-like particles, decays directly into hadronic final states will become dominant.  In the general case \cite{Bauer:2017ris}, the decay mode $\eta'_d \rightarrow \pi \pi \pi$ would be significant.  However, this decay rate depends on the coupling of the ALP to gluons, which is zero for $\eta'_d$, and on a difference between couplings to up and down quarks which is also zero.  Thus, we have $\Gamma(\eta'_d \rightarrow \pi \pi \pi) \approx 0$.  Other direct hadronic decays are also dominated by the ALP-gluon coupling, so we will neglect such decays in this work and only include decays to heavy quarks (charm and bottom.)

\subsection{Dark glueballs and other mesons} 
\label{subsec:glueballs}

In the limit that the single light flavor is 
somewhat heavy (i.e., $\Lambda_d \ll m_{n,0} \ll m_{\rm eq}$), the lightest meson state will be 
the lightest $0^{++}$ glueball.  
The early universe cosmology of pure glue theories 
has been analyzed in 
Ref.~\cite{Forestell:2016qhc,Forestell:2017wov}, 
examining the relic abundances of the glueball states.  
Their analysis suggests that heavier glueballs efficiently downscatter into lighter glueballs, with the dominant abundance of relic glueballs in the $0^{++}$ state.  

In the opposite limit $m_{n,0} \lesssim \Lambda_d$, 
it is expected that the $\eta'$, the $\sigma$,
and the lightest glueball meson could be present,
with mixing among the states with the same $J^{PC}$.
Without additional nonperturbative input, 
we do not know which state is the lightest, 
and more importantly, the relative abundance of these
light states.  However, what we can do is calculate
the lifetimes of these states \emph{in isolation}, 
providing a set of sufficient conditions for the parameter
space of HSDM model to satisfy to ensure
there are no long-lived mesons whose decays disrupt
BBN\@.

\subsection{Lightest $0^{++}$ glueball}

\begin{figure*}
\begin{center}
\includegraphics[width=0.49\textwidth]{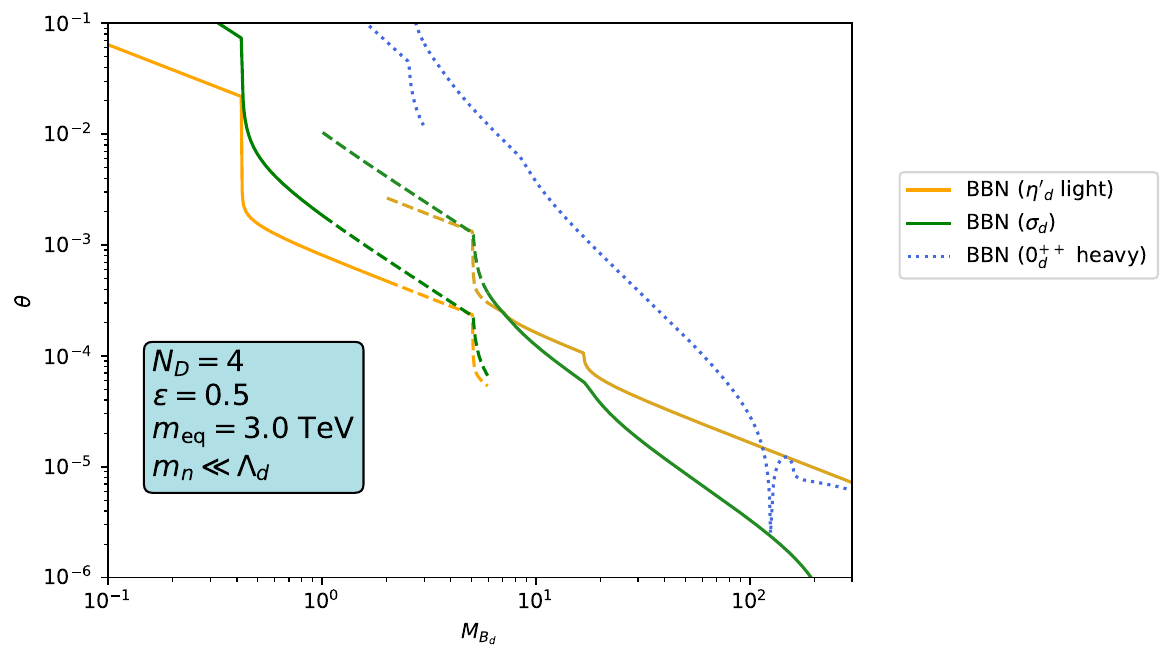}
\includegraphics[width=0.49\textwidth]{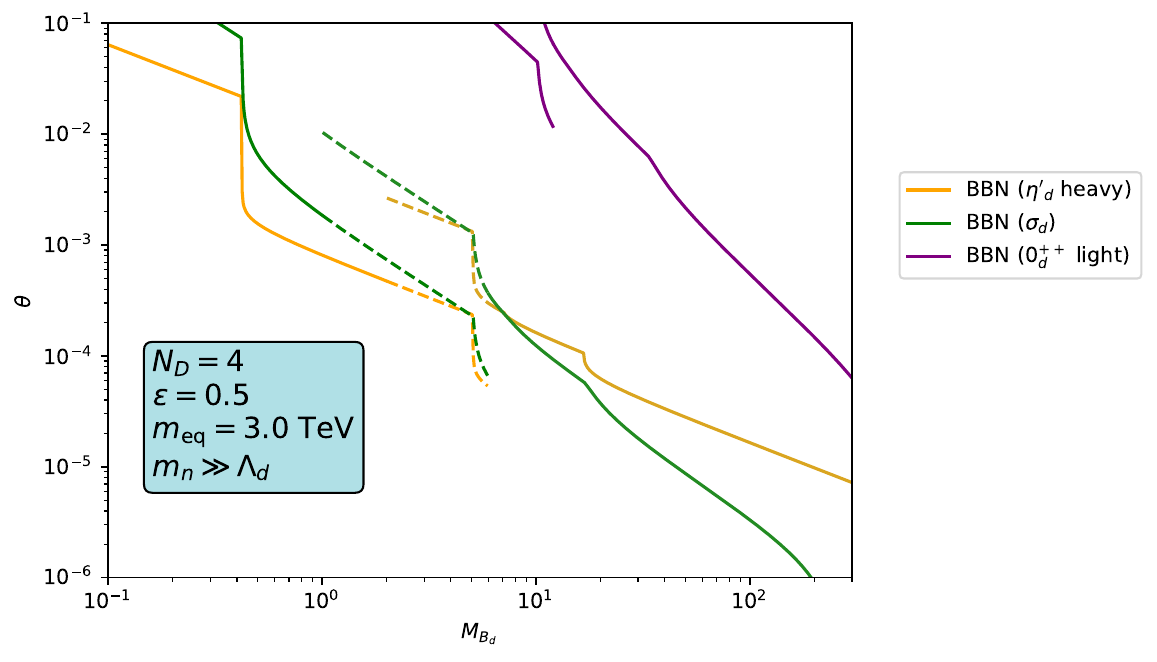}
\end{center}
\caption{Collection of BBN bounds from \cref{fig:BBN_eta} and \cref{fig:BBN_scalar} for all meson states, in the light-quark limit $m_n \ll \Lambda_d$ (left) and the heavy-quark limit $m_n \gg \Lambda_d$ (right).  The comparison is done within the dark equilibration sector UV completion, so that all bounds can be shown on a common set of axes, with the parameter $\theta$ versus the dark baryon mass $M_{B_d}$.}
\label{fig:BBN_heavy_light}
\end{figure*}

Here we provide the lifetime of the lightest $0^{++}$ glueball meson in the limit that it does not mix 
with any other meson state.  This is physically 
realized if the single light flavor is somewhat heavy, 
so that nonperturbative sector is pure glue yielding a spectrum of mesons that are dominantly just glueballs.
Lattice studies \cite{
Morningstar:1999rf,Chen:2005mg,Athenodorou:2021qvs}.
have calculated the mass spectrum of the glueballs in $SU(3)$ and some $SU(N_c)$ theories, finding that the lightest glueball is the $0^{++}$.  

Our purpose is to calculate the lifetime of the lightest $0^{++}$ meson state in the limit that all of the dark quarks are heavy.  The $0^{++}$ state decay can be estimated by assuming it is dominated by the dimension-6 operator, ${\rm Tr}[G_{\mu\nu} G^{\mu\nu}] H^\dagger H$, given in \cref{eq:Lint_EW_symmetric}. In the regime $m_{0^{++}} < 2 m_h$, we can use the the results from \cite{Juknevich:2009ji,Juknevich:2009gg,Batz:2023zef} to obtain
\begin{equation}
  \Gamma_{0^{++} \rightarrow \xi\xi} \, = \, \frac{c_G^2 v^4}{\meq^4} \left[ \frac{ \mathbf{F^{S}_{0^{+}}}}{v (m_h^2 - M_{0^{++}_d}^2)} \right]^2 \Gamma^{\rm SM}_{h \rightarrow \xi\xi}(M_{0^{++}_d}^2) \, ,
\end{equation}
the width of the $0^{++}$ glueball into the SM final states $\xi\xi$ (where $\xi$ is any SM state that a Higgs boson with a mass of $M_{0^{++}_d}$ that could decay into).  Over the range of masses that we focus on in this paper, the decay width will be largely dominated by $\bar{f} f$ quark or lepton pairs, with decays to $WW$ becoming relevant near the Higgs mass. The glueball width is expressed in terms of the Higgs boson width
$\Gamma^{\rm SM}_{h \rightarrow \xi\xi}(M_{0^{++}_d}^2)$ where the Higgs mass is replaced with the mass of the $0^{++}$; we adopt formulas from \cite{Marciano:1993sh,Dawson:2017ksx} for the Higgs width.  The annihilation matrix element of the $0^{++}$ state is expressed as $\langle 0|S|0^{++}\rangle \equiv \mathbf{F^{S}_{0^{+}}}$ in terms of the non-perturbative decay constant $\mathbf{F^{S}_{0^{+}}}$ with mass dimension three.  The coefficient of the effective interaction arises from integrating out the equilibration sector.  We estimate
\begin{eqnarray}
    \frac{c_G v^2}{\meq^2} &\simeq& \frac{\alpha_d \, y_{ln} y_{ln}' v^2}{3 \pi m_{\rm eq}^2}  \, , 
\end{eqnarray}
arising from one-loop box diagram contributions from the equilibration sector fermions\footnote{There is also a contribution from simultaneously light and heavy dark quarks in the same loop, but the loop contribution remains dominated by the heavy dark quark scale of the equilibration sector.} and $\alpha_d = g_d^2/(4\pi)$ in terms of the dark sector coupling constant $g_d$.
Using eqs.~(\ref{eq:yln}),(\ref{eq:ylnp}), and neglecting contributions of order $\epsilon^2$, this becomes
\begin{eqnarray}
    \frac{c_G v^2}{\meq^2} &\simeq& \frac{2 \alpha_d}{3 \pi} \theta^2 \, . 
\end{eqnarray}
Although it cancels in the prediction of the physical decay width below, we adopt $\alpha_d \sim 0.1$ for numerical conversion between $c_G$ and $\theta$.  The full decay rate is therefore given by
\begin{eqnarray}
\Gamma_{0^{++},\rm tot} &=& \sum_\xi \; \Gamma(0^{++} \rightarrow \xi\xi) \\
&=& \frac{4 \alpha_d^2}{9 \pi^2} \theta^4
\left[ \frac{ \mathbf{F^{S}_{0^{++}}}}{v (m_h^2 - M_{0^{++}_d}^2)} \right]^2
\Gamma^{\rm SM}_{h,{\rm tot}}(M_{0^{++}_d}^2) \, . \nonumber
\end{eqnarray}
For $SU(3)$, lattice results found the decay constant for the $0^{++}$ state to be \cite{Chen:2005mg,Juknevich:2009gg}
$4 \pi \alpha_d {\bf F_{0^{++}}} = 3.1 M_{0^{++}_d}^3$.  In the large-$N_D$ limit, this decay constant is expected to scale as $N_D^1$ \cite{Giacosa:2024scx}, so we have
\begin{equation}
\mathbf{F_{0^{++}}^{S}} = \frac{3.1}{4\pi \alpha_d} \frac{N_D}{3} M_{0^{++}_d}^3.
\end{equation}
We can substitute this this relation into the total width to obtain 
\begin{equation}
\Gamma_{0^{++},\rm tot} \;=\; \frac{(3.1)^2}{36 \pi^4} \left(\frac{N_D}{3}\right)^2 \theta^4 \frac{M_{0^{++}_d}^6}{v^2 (m_h^2 - M_{0^{++}_d}^2)^2}
\Gamma^{\rm SM}_{h,{\rm tot}}(M_{0^{++}_d}^2) \, .
\end{equation}
We show the bounds on the parameter using from the glueball $0^{++}$ decay in \cref{fig:BBN_scalar}(left).

We can do an analytic comparison to $\eta'$ decay under several assmptions: i) both $\eta'$ and $0^{++}$ decay are dominated by just one channel (e.g.\ to $b\bar{b}$), and ii) set their masses equal $m_{\eta'} = m_{0^{++}}$. With these approximations, we obtain
\begin{equation}
\frac{\Gamma_{0^{++}}}{\Gamma_{\eta'}} \; \simeq \; \frac{1}{12 \pi^4} \left( \frac{m_{0^{++}}^2}{m_h^2 - m_{0^{++}}^2} \right)^2  \left( 1 - \frac{4 m_f^2}{m^2_{0^{++}}} \right) \frac{N_D}{\epsilon^4}
\end{equation}

\subsection{$\sigma$ meson}

Finally, we consider the $\sigma$ meson state formed from 
$\bar{q}q$ with $J^{PC} = 0^{++}$.
Here we calculate the lifetime of the $\sigma$ meson
assuming there is no mixing with other states.
The lifetime calculation follows closely with
the decay rate of the $0^{++}$ glueball, 
except that the operator leading to the decay
is dimension-5, $H^\dagger H \bar{\Psi} \Psi$.
The coefficient of this operator is given by
$c_s$ in \cref{eq:cs_theta}.  Putting this
all together, we obtain
\begin{equation}
    \Gamma_{\sigma \rightarrow \xi\xi} \;=\;
    4 \, \theta^4 \left( \frac{m_{\rm eq}}{v} \right)^2
    \left( \frac{\mathbf{F}_\sigma}{m_h^2 - m_\sigma^2} \right)^2
\Gamma^{\rm SM}_{h \rightarrow \xi\xi}(m_\sigma^2) \, ,
\end{equation}
the width of the $\sigma$ meson state into the SM final states $\xi\xi$ (where again $\xi$ is any SM state that a Higgs boson with a mass of $m_\sigma$ that could decay into). Like the $0^{++}$ glueball width, the $\sigma$ width is expressed in terms of the Higgs boson width
$\Gamma^{\rm SM}_{h \rightarrow \xi\xi}(m_\sigma^2)$ where the Higgs mass is replaced with $m_\sigma$.  The annihilation matrix element of the $\sigma$ state is expressed in terms of $\langle 0|\bar{\Psi}\Psi|\sigma\rangle \equiv \mathbf{F_\sigma}$ in terms of the non-perturbative decay constant $\mathbf{F_\sigma}$ with mass dimension two.

The large-$N_D$ scaling of $\mathbf{F}_{\sigma}$, as with any other meson decay constant \cite{Giacosa:2024scx}, is $\mathbf{F}_{\sigma} \sim N_D^{1/2}$.  We are not aware of any lattice calculations of the magnitude of the scalar decay constant; we will rewrite it as 
\begin{equation}
\mathbf{F}_{\sigma} \, = \, \sqrt{\frac{N_D}{3}} \chi_{\sigma} m_{\sigma}^2,
\end{equation}
 where $\chi_{\sigma}$ is an $\mathcal{O}(1)$ parameter that we will set equal to 1 in our numerical results below.
Like for the previous set of meson states, BBN bounds on the parameter space can be obtained and are shown in 
\cref{fig:BBN_scalar}(right).

We can again take the ratio of this width to $\Gamma_{\eta'}$, using the same approximations made above for the $0^{++}$, i.e., $m_{\eta'} = m_\sigma$ and assuming the decay rates are dominated by just one fermionic channel,
and we obtain
\begin{eqnarray}
\frac{\Gamma_\sigma}{\Gamma_{\eta'}}  &=&  \frac{1}{N_D \epsilon^4} \left( \frac{m_{\rm eq}}{m_\sigma} \right)^2
\left( \frac{\mathbf{F_\sigma}}{m_\sigma f_{\eta'}} \right)^2
\left( \frac{m_\sigma^2}{m_h^2 - m_\sigma^2} \right)^2 \nonumber \\
& & \qquad \times \left( 1 - \frac{4 m_f^2}{m_\sigma^2} \right) \\
&=& \frac{1}{\epsilon^4} \left( \frac{m_{\rm eq}}{m_\sigma} \right)^2
\left( \frac{ \chi_{\sigma} m_{\sigma}}{f_{\eta'}}  \right)^2
\left( \frac{m_\sigma^2}{m_h^2 - m_\sigma^2} \right)^2  \nonumber \\
& & \qquad \times 
\left( 1 - \frac{4 m_f^2}{m_\sigma^2} \right).
\end{eqnarray}
Whether $\Gamma_\sigma$ is larger or 
smaller than $\Gamma_{\eta'}$
depends on the relative size of the factors  $(1/\epsilon^4) \times (m_{\rm eq}/m_\sigma)^2$ 
(that suggests $\Gamma_\sigma$ should be larger)
versus the Higgs exchange propagator squared $(m_{\sigma}^2/(m_h^2-m_{\sigma}^2))^2$
(that suggests $\Gamma_\sigma$ should
be suppressed for $m_\sigma \ll m_h$).

\subsection{Summary of meson decay bounds}

Putting all of this together, in  \cref{fig:BBN_heavy_light}, we summarize the BBN bounds on the $\eta'$, $0^{++}$ glueball, and the $\sigma$ meson in the 
light quark ($m_n \ll \Lambda_d$)
and heavy quark ($m_n \gg \Lambda_d$)
regimes using the dark baryon mass
to set the common set of scales.
We see that ensuring the $\eta'$ 
lifetime satisfies the BBN constraints
dominates the parameter space restrictions
in much of 
the light quark regime, 
while the ensuring the $0^{++}$ 
lifetime satisfies the BBN constraints
determines the parameter space 
restrictions in the heavy quark regime.

\section{\label{sec:abundance} Cosmology and Relic Abundance}

Having emphasized the broad range of phenomenology that is possible in
the model, one of the unanswered questions in this work is the
cosmological evolution of HSDM and the mechanism for producing the
cosmological abundance of dark baryons (and/or dark anti-baryons).

In general, the presence of equilibration sector fermions with
electroweak interactions means that the dark sector will be in thermal
equilibrium with the Standard Model for temperatures $T \gtrsim
\meq$ with $\meq \sim$ 3 TeV or higher.  As the temperature drops
below $\meq$, the equilibration-sector heavy fermions will annihilate
or decay efficiently down to the light dark sector fermions. 
In the regime where the light dark sector fermions are not parametrically heavier
than the confinement scale, dark confinement 
of the dark-quark-gluon plasma 
into dark hadrons will occur first,
once the temperature drops below the dark confinement scale.
Below confinement, we anticipate
the spectrum is 
made of the lightest baryon (the dark matter particle) and the light
dark mesons.  The ratio of the
dark baryon mass to the dark $\eta_d'$
mass is expected to be $N_D/2$,
where $N_D \geq 3$, as we  
discussed in section \ref{sec:confinedND}
(see equation \ref{eq:quark_ratio}).
Both the dark baryon and mesons become 
non-relativistic 
in the confined phase. For a brief
period there will be a bath of strongly interacting mesons and baryons.
As the temperature continues to drop,
it is energetically favorable
for dark baryons and anti-baryons to 
annihilate into dark $\eta'$s,
leading to a freeze-out abundance
of dark baryons once the Hubble rate
exceeds the rate for these interactions.
The rate of this conversion strongly depends on the dark baryon--dark anti-baryon annihilation
cross section (see below). 
The remaining $\eta'_d$s decay 
into the SM, discussed in detail 
in section \ref{subsec:eta_decay}.
This decay will transfer entropy from 
the dark sector to the SM, 
completing by the time of BBN 
(as demanded by the constraint 
that the $\eta_d'$ decay occurs
before the onset of BBN).

The dark baryon--anti-baryon annihilation cross section into dark mesons is intrinsically nonperturbative, but we can take guidance from expectations at small and large $N_c$.
For small $N_c$, we know from QCD 
($N_c=3$) that the baryon--anti-baryon
cross section is large, leading to the highly efficient annihilation of the symmetric abundance in the early 
universe.
If HSDM with small $N_c$ followed this
estimate, we anticipate 
\emph{underproducing} dark
matter for all of the mass scales considered in this paper.  (If there were an asymmetric abundance of dark baryons, this will efficiently eliminate the symmetric component leaving only the 
asymmetric one, just as happens
in the SM). 
On the other hand, at large $N_c$,
Witten pointed out that baryon--anti-baryon annihilation is
exponentially suppressed 
\cite{Witten:1979kh} (see also
\cite{Morrison:2020yeg}). 
In this scenario, we expect a
very small cross section that would cause
an \emph{overproduction} of dark matter 
at freeze out (that would overclose the
universe). 
A large range of dark matter abundances is therefore expected in our theory, given the range of mass scales and the exponential sensitivity to $N_c$. 
Nevertheless, we anticipate that there is
some discrete set of $N_c$
that, combined with the relevant 
mass scales ($\Lambda_d$, $m_d$),
could give rise to a symmetric 
abundance of dark baryons and
anti-baryons that matches the
cosmological abundance of dark matter
in the universe.  We cannot calculate
the precise value or range of parameters, since this requires 
a nonperturbative calculation of the
baryon--anti-baryon annihilation
cross section to mesons.
Estimates for $N_c$ can be gleaned 
from a similar one-flavor theory, 
where the freeze-in abundance of 
dark baryons and anti-baryons
matched the cosmological abundance when $N_c \approx 10$ \cite{Morrison:2020yeg}.

Careful readers will note that the quantum numbers given in Table~\ref{tab:HSDM-charges} lead to a vector-like fermion mass spectrum that has one equilibration sector fermion with charge $-1$.  By itself this is not a problem, since there are anti-fermions with exactly the opposite electric charges (and opposite baryon numbers), and so if both fermions and anti-fermions are equally populated in the early universe, there is no cosmological constraint from requiring the vanishing of total electric charge of the dark sector.  This occurs naturally with symmetric abundance mechanisms that populate an equal amount of dark baryons and dark anti-baryons. 

Another way to achieve the cosmological abundance would be through an asymmetric mechanism, however the correlation between electric charge and baryon number suggests this would require a careful examination of how the model could be extended to obtain an asymmetry in dark baryon number while also maintaining the electric charge neutrality of the dark sector.  One way to ensure electric charge neutrality solely within the baryon sector (and also electric neutrality within the anti-baryon sector) would be to extend HSDM to a theory with more flavors of dark quarks transforming under the electroweak group.  For example, one could add a fourth flavor to the model that is a singlet under $SU(2)_L$ with  hypercharge $+1$, and thus electric charge $+1$, compensating for the $-1$ charge from the doublet.
We leave further investigations of an asymmetric mechanism to future work.

We would be remiss to not emphasize that there are also a vast set of indirect detection constraints 
\cite{Slatyer:2017sev,Leane:2020liq,Slatyer:2021qgc} that can constrain a large range of dark matter masses
in the case where there is a symmetric abundance of dark baryons and anti-baryon.
In particular, the parameter space identified in Fig.~\ref{fig:bounds} still requires careful examination with respect to the plethora of indirect astrophysical constraints.  A detailed analysis of the annihilation rates of dark baryons and anti-baryons into the SM requires nonperturbative information, and so we do not have estimates to present here.  There are also potential constraints on asymmetric dark matter that can 
accumulate inside neutron stars 
\cite{Bertoni:2013bsa,McKeen:2018xwc,Gresham:2018anj,Garani:2018kkd}.
These bounds do have a requirement that dark matter scatters off the nuclei within the stars, losing energy, and accumulating inside the core of the star,
which would need careful re-examination for HSDM\@.

\section{\label{sec:conclusion} Conclusions and Outlook}

\begin{figure}
\begin{center}
\includegraphics[width=0.95\textwidth]{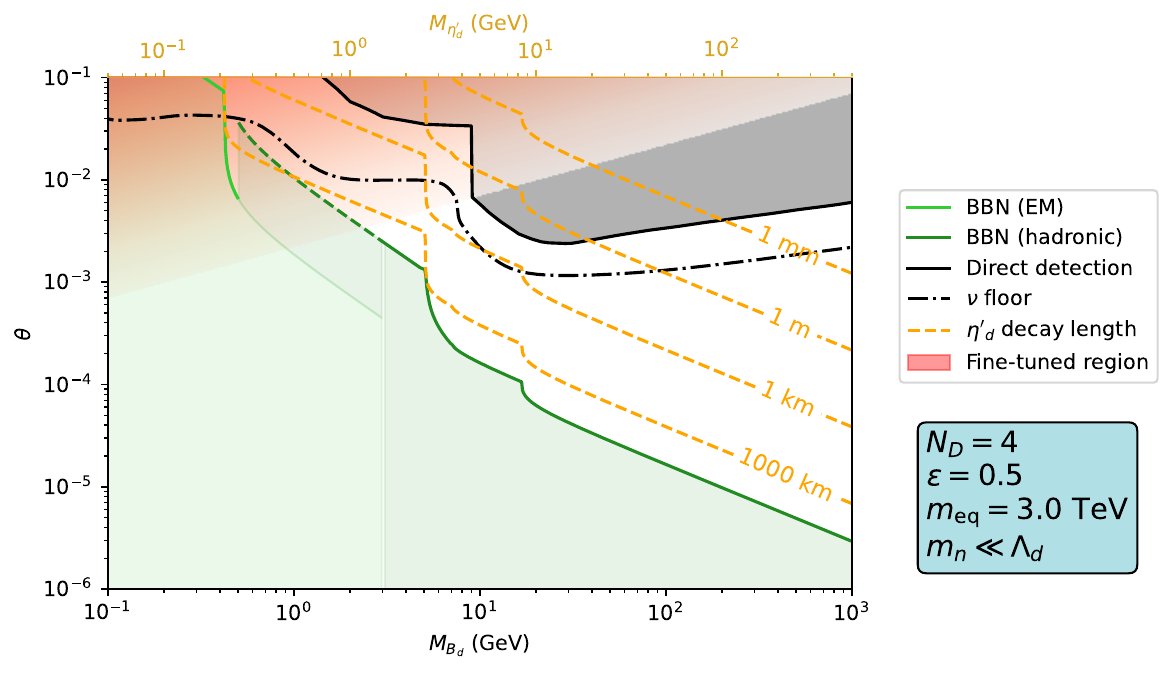}
\includegraphics[width=0.95\textwidth]{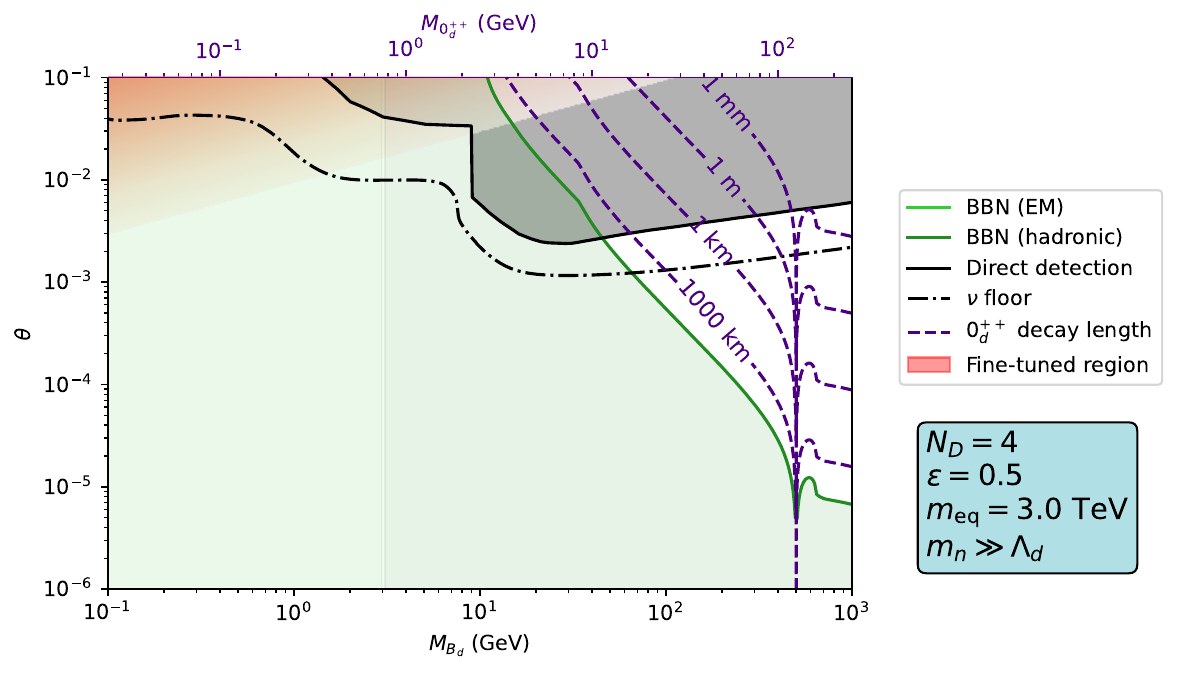}
\end{center}
\caption{Top: The allowed region of the HSDM mass and parameter
  $\theta$ in the light quark limit, $m_n \ll \Lambda_d$. As shown in the plot, other model parameters are fixed to the fiducial values $N_D = 4$, $\epsilon = 0.9$, and $\Lambda = 5$ TeV.  Bounds shown include direct detection through Z exchange (see \cref{sec:direct}), BBN bounds on $\eta'$ decays (\cref{subsec:eta_decay}), and a disfavored region with fine-tuning (\cref{subsec:ewsbfinetuning}).  For the BBN bounds, the solid/dashed curve is at the fiducial mass ratio $M_{B_d} / M_{\eta'_d} = N_D/2$; the nearby dotted lines show the effect of varying this ratio over the full range discussed in \cref{sec:confinedND}.  The upper triangular region shaded in red shows the parameter space disfavored by fine-tuning, see \cref{subsec:ewsbfinetuning}.    Note that the overall lower bound for the HSDM mass is in the few GeV region.
  Bottom: Same as top plot except that the parameter region
  is in the heavy quark limit, $m_n \gg \Lambda_d$.
} 
\label{fig:bounds}
\end{figure}

We have developed a theory of light composite dark matter, called Hyper Stealth Dark Matter, 
that equilibrates with the SM through \emph{just} SM interactions, and is consistent with
current experimental constraints down to a few GeV\@.
We have taken some benchmark values
for the parameters of the theory,
namely $N_D = 4$ dark colors,
$m_{\rm eq} = 3$~TeV (close to
the anticipated 
LHC bounds on dark meson production),
and the difference between the 
Yukawa couplings taken to be order one
($\epsilon = 0.5$).
Our final results are shown 
in the light quark limit
$m_n \ll \Lambda_d$ in Fig.~\ref{fig:bounds}(top)
where we find the dark baryon can be
as light as a few GeV without 
fine-tuning parameters of the model.
In the heavy quark limit $m_n \gg \Lambda_d$, 
shown in Fig.~\ref{fig:bounds}(bottom),
we find the dark baryon can be as light as 
about $50$~GeV\@. 
The lower bounds arise
from the intersection of two competing requirements: i) the equilibration sector of the model must
be sufficiently heavy, at least several TeV, to avoid bounds on the heavy meson sector from colliders, and ii) the lightest
dark meson must decay before BBN\@. 
Raising the scale of the equilibration sector 
causes several consequences:  the 
light dark meson lifetimes increase,
the elastic scattering cross section for 
direct detection decreases, and 
tuning among the UV parameters to obtain a light fermion mass is increased.

\begin{figure}
\begin{center}
\includegraphics[width=0.95\textwidth]{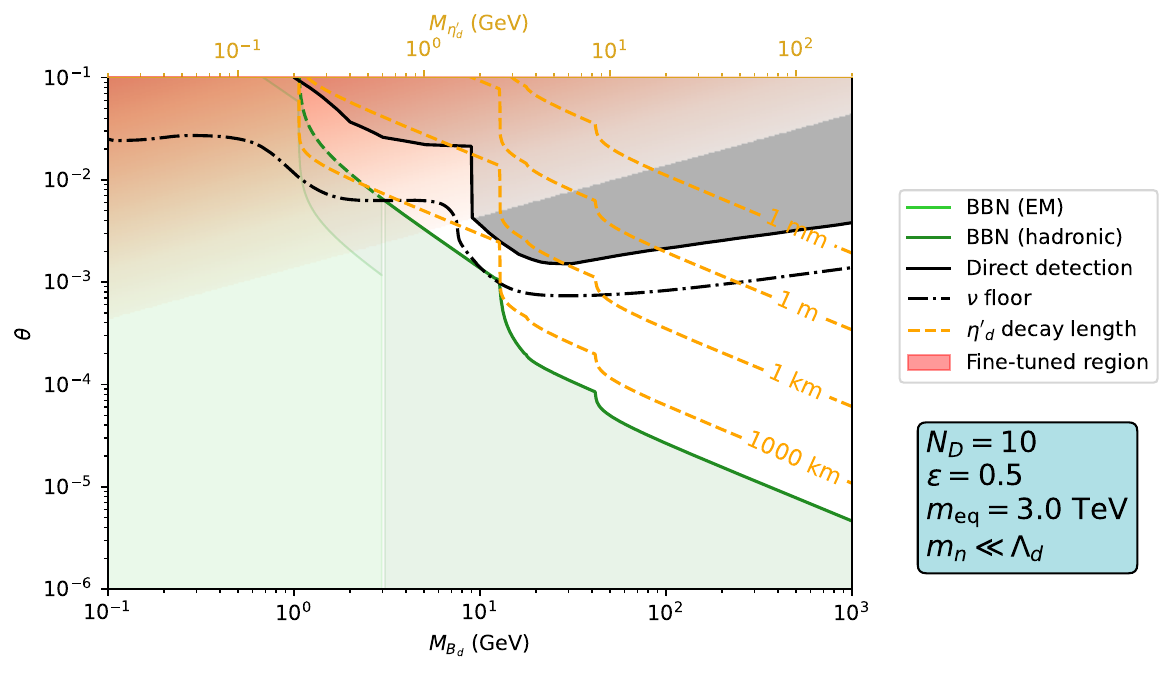}
\includegraphics[width=0.95\textwidth]{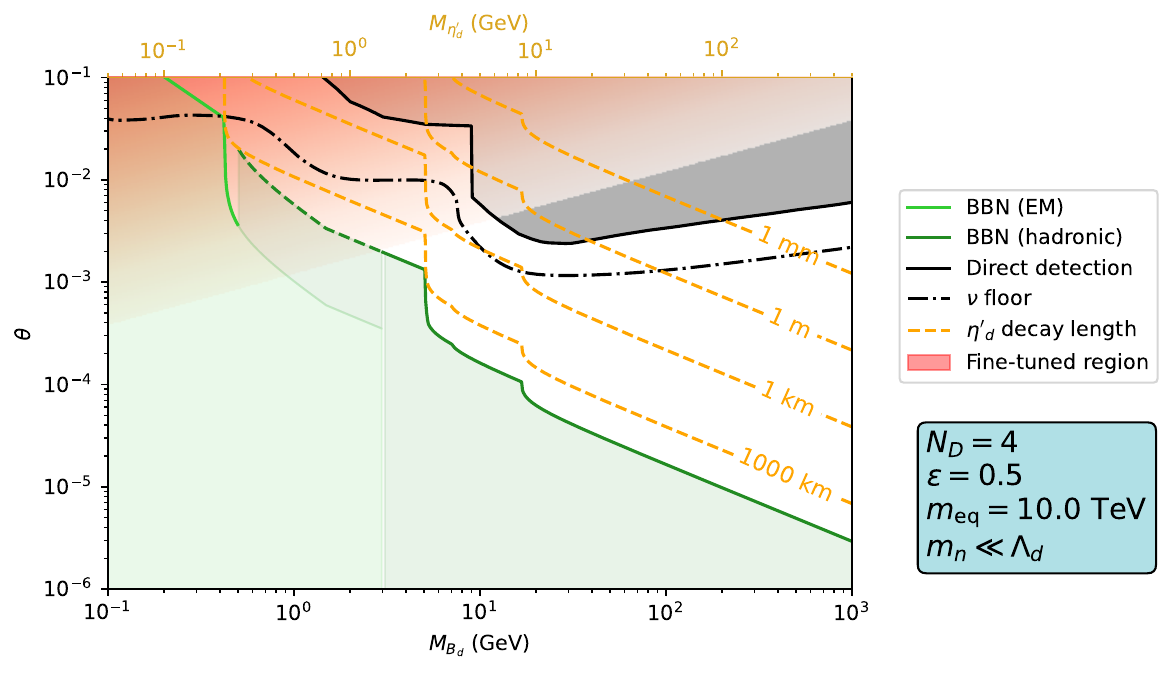}
\end{center}
\caption{Similar to \cref{fig:bounds} (top), also in the light quark limit $m_n \ll \Lambda_d$: Alternative constraint plots varying the heavy scale $\meq$ and the number of dark colors $N_D$.  Note that increasing $N_D$ also indirectly results in stronger BBN bounds at light $M_{B_d}$ by increasing the ratio $M_{B_d} / M_{\eta'_d}$.
} 
\label{fig:bound_variants}
\end{figure}

We have also explored modest variations
of the parameter space 
in the light quark regime in \cref{fig:bound_variants}. 
In \cref{fig:bound_variants}(top),
we show how the parameter space 
changes when increasing the number
of dark colors to $N_D = 10$,
while in
\cref{fig:bound_variants}(bottom)
we show the changes when 
$m_{\rm eq} = 10$~TeV\@.
In both cases, while there are 
observable differences from 
shown in \cref{fig:bounds}(top), 
the changes are modest, 
demonstrating the robustness of our
results with respect to the 
underlying parameters of the theory.

The opening up of composite dark matter between the few GeV to hundreds of GeV scale,
as shown in figs.~\ref{fig:bounds},\ref{fig:bound_variants} 
is fascinating for a variety of reasons:  
\begin{itemize}
\item[i)]  Direct detection off nuclei is possible throughout the mass range, though the rates may be much smaller than anticipated future direct detection sensitivity; 
\item[ii)]  The lightest meson is anticipated to have a long lifetime, motivating searches at both the 
LHC \cite{Alimena:2019zri} and 
dedicated long-lived particle facilities 
such as 
FASER \cite{FASER:2018eoc} and MATHUSLA
\cite{MATHUSLA:2018bqv,MATHUSLA:2020uve} (see also \cite{Cheng:2019yai} for detailed collider studies of one-flavor meson decays in the context of a different UV completion);
\item[iii)]  Self-interactions of dark baryons are expected to be below existing bounds from galaxy cluster mergers, motivating continued investigations of large scale structure that include the effects of (small) self-interactions; 
\item[iv)]  If the model could be fused with a viable asymmetric dark matter production mechanism (though an extension of the model) that populated dark baryons over dark anti-baryons, the lighter dark matter mass scales permit the number densities of dark matter to be comparable to SM baryons;
and finally, 
\item[v)] the $SU(N)$ gauge theory with one light flavor will have a phase transition at temperatures near the confinement scale
that may be first order for some $N$ and some range of the lightest dark quark masses.  If the transition is first order, we expect a stochastic background of gravitational waves \cite{Schwaller:2015tja} may be detectable at future gravitational observatories. 
\end{itemize}

The equilibration sector also provides collider physics opportunities for detection.  Current or future proposed collider experiments with energies that can
probe up to the $10$ TeV region for new particles have the opportunity to see the rich electroweak-charged heavy spectrum of HSDM.

Finally, lattice simulations can also provide very useful guidance to understand the structure and parameter space of the theory.  Among the highest priorities would be:  the spectrum of the light mesons and baryons as a function of the light fermion mass; the meson decay constants; the order of the phase transition, latest heat, and other observables relevant to the stochastic gravitational wave signal; the dark baryon self-interaction rates as a function of the light fermion mass; etc.



\onecolumngrid

\pagebreak

\twocolumngrid

\section*{Acknowledgments}

We thank David Curtin, Sally Dawson, and Tom DeGrand for useful
discussions.  G.D.K.\ thanks the Mainz Institute for Theoretical
Physics and the CERN Department of Theoretical Physics where part of
this work was completed.  This work was also produced by the FermiForward
Discovery Group (G.T.F.), LLC under Contract No. 89243024CSC000002
with the U.S. Department of Energy, Office of Science, Office of High
Energy Physics.  The work of G.D.K.\ is supported in part by the U.S.\
Department of Energy under grant number DE-SC0011640.  The work of
E.N.\ is supported by the U.S.\ Department of Energy under Grant
Contract DE-SC0010005. D.S. was supported by UK Research and
Innovation Future Leader Fellowship MR/S015418/1 \& MR/X015157/1 and
STFC grants ST/T000988/1 \& ST/X000699/1.  This work (P.M.V.) was performed
under the auspices of the U.S. Department of Energy by Lawrence
Livermore National Laboratory under Contract DE- AC52-07NA27344.

\appendix

\section{Generalized trace anomaly Higgs coupling \label{app:trace_anomaly}}

 In this appendix, we revisit the calculation of \cite{Shifman:1978zn}, and adapt their results slightly to cases with arbitrary numbers of colors $N_D$ and heavy quark flavors. The trace anomaly is equal to
\begin{equation}
\tr[\theta] \, = \, \frac{\beta(\alpha_D)}{4\alpha_D} \tr [GG] + m_n \bar{n} n + \sum_{\lambda} m_\lambda \bar{\lambda} \lambda \, ,
\end{equation}
where we have a single light fermion $n$, and $\lambda$ labels the $N_\lambda$ heavy mass eigenstates.  $\beta(\alpha_D)$ is the one-loop Yang-Mills beta function, which for general numbers of colors $N_D$ and flavors $N_f$ is
\begin{equation}
\beta(\alpha_D) \, = \, -\frac{\alpha_D^2}{2\pi} \left[ \frac{11}{3} N_D - \frac{2}{3} (N_\lambda + 1) \right] \, ,
\end{equation}
where $\alpha_D \equiv g_D^2 / (4\pi)$ and $\beta(\alpha_D) \equiv d\alpha_D / d(\log \mu)$, with $\mu$ the RG scale.

Now, consider a baryon $B_d$ composed of the light $n$ fermions.  The matrix element of the trace anomaly with the baryon $B_d$ gives the baryon mass operator:
\begin{equation}
\langle B_d | \, \tr[\theta] \, | B_d \rangle \, = \, \mdm \bar{B}_d B_d \, .
\end{equation}

Next, we integrate out the heavy fermions; up to corrections of order $1/m_\lambda^2$, the only effect is from a triangle diagram involving the heavy quarks (see \cite{Shifman:1978zn} for a sketch and further details.)  Calculation of the diagram in $\overline{\rm MS}$ yields the amplitude
\begin{equation}
\mathcal{M}_{\rm triangle} \, \sim \, -\tr[T^a_{R} T^a_{R}] \frac{8}{3} \frac{\alpha_D}{4\pi} \, ,
\end{equation}
where since our fermions are in the fundamental representation, the color trace $\tr[T^a T^a] = 1/2$.  Integrating the heavy quarks out provides a contribution proportional to the gluon kinetic term $(-1/4 \tr[GG])$, which means that this diagram implies the replacement (up to corrections of order $1/m_\lambda^2$)
\begin{equation}
\sum_\lambda m_\lambda \bar{\lambda} \lambda \; \rightarrow \; -\frac{2}{3} \frac{\alpha_D}{8\pi}  N_\lambda \tr[GG] \, .
\end{equation}
As a check, this matches the reference \cite{Shifman:1978zn} when used for the example of QCD.  This replacement can be substituted into the trace anomaly expression as in the reference, yielding
\begin{equation}
\tr[\theta] \, = \, \frac{\tilde{\beta}(\alpha_D)}{4\alpha_D} \tr [GG] + m_n \bar{n} n \, ,
\end{equation}
where $\tilde{\beta}(\alpha_D)$ is the $\beta$-function from above but with $N_\lambda = 0$ (only keeping the contribution from the one light fermion $n$),
\begin{equation}
\tilde{\beta}(\alpha_D) \, = \, -\frac{\alpha_D^2}{6\pi} (11 N_D - 2) \, .
\end{equation}
Now, our objective is to obtain the coupling of the Higgs to the baryon.  In the Standard Model, because the masses of quarks arise entirely from Yukawa couplings, the Higgs boson coupling is directly proportional to the mass.  However, in general (and specifically in the Hyper Stealth Dark Matter model), only a fraction of the total mass $m_\lambda$ comes from the Higgs.  Defining $\kappa_\lambda \equiv (y_\lambda v) / m_\lambda$, we have for the Higgs-heavy fermion interaction
\begin{equation}
\mathcal{L}_h \, = \, h \sum_\lambda  y_{\lambda} \bar{\lambda} \lambda \, = \, h \sum_\lambda \kappa_\lambda \frac{m_\lambda}{v} \bar{\lambda} \lambda \, .
\end{equation}

Finally, we can compute the coupling of the Higgs to the light baryon induced by the heavy-quark current, assuming $\kappa_\lambda$ is constant:
\begin{eqnarray}
\lefteqn{h \langle B_d | \sum_\lambda y_\lambda \bar{\lambda} \lambda | B_d \rangle} & & \nonumber \\
&=& -\frac{N_\lambda \kappa_\lambda \alpha_D}{12\pi v} h \langle B_d | \, \tr [GG] \, | B_d \rangle \nonumber \\
&=& -\frac{N_\lambda \kappa_\lambda \alpha_D}{12\pi v} h \langle B_d | \frac{4\alpha_D}{\tilde{\beta}(\alpha_D)} \left( \tr[\theta]  - m_n \bar{n} n \right) | B_d \rangle \nonumber \\
&=& \frac{N_\lambda \kappa_\lambda}{12\pi v}  \frac{24\pi}{11 N_D - 2} h \left( \mdm \bar{B}_d B_d - \langle B_d | m_n \bar{n} n | B_d \rangle \right) \nonumber \\
&=& \frac{2N_\lambda \kappa_\lambda}{v} \frac{1}{11N_D - 2} h ( \mdm - \langle B_d | m_n \bar{n} n | B_d \rangle ) \nonumber \\
&=& \frac{2N_\lambda \kappa_\lambda}{11N_D - 2} \frac{\mdm}{v} \left(1 - f_n^{(B_d)} \right) \, ,
\end{eqnarray}
using the definition of the light-fermion sigma term \cref{eq:sigma_n}.

The total Higgs coupling to the baryon can be obtained by adding in the light fermion current as well:
\begin{equation}
g_{B_d,h} h \bar{B_d} B_d \, = \, h \langle B_d | y_n \bar{n} n + \sum_\lambda y_\lambda \bar{\lambda}\lambda | B_d \rangle \, .
\end{equation}
For the light fermion, we can rewrite
\begin{eqnarray}
\langle B_d|y_n \bar{n} n | B_d \rangle &=& \frac{\kappa_n}{v} \langle B_d | m_n \bar{n} n | B_d \rangle \nonumber \\
&=& \kappa_n \frac{\mdm}{v} f_n^{(B_d)},
\end{eqnarray}
with $\kappa_n \equiv (y_n v) / m_n$.
Thus, the total Higgs-baryon coupling is
\begin{equation}
g_{B_d,h} \, = \, \frac{\mdm}{v} \kappa_n f_n^{(B_d)} + \frac{2N_\lambda \kappa_\lambda}{11N_D - 2} \frac{\mdm}{v} \left(1 - f_n^{(B_d)} \right) .
\end{equation}

Now, we specialize to the HSDM model, which means that $N_\lambda = 2$.  Moreover, within this model we can compute the $\kappa$ factors.  From \cref{eq:nZ_current}, we have
\begin{eqnarray}
y_n &=& \frac{c_s v}{2\Lambda} \\
\Rightarrow \; \; \kappa_n &=& \frac{y_n v}{m_n} \, = \, \frac{c_s v^2 / (2\Lambda)}{m_{n,0} - c_s v^2 / (2\Lambda)} \\
&=& \frac{\theta^2 \Lambda}{m_{n,0} - \theta^2 \Lambda} \, = \, \frac{\theta^2 \Lambda}{m_n} \, .
\end{eqnarray}
In this form, the value of $\kappa_n$ is clearly related to the fraction of the light fermion mass which arises from the Higgs mechanism versus the vector mass.  Similarly, for the heavy fermions, we find
\begin{eqnarray}
\kappa_\lambda &=& \frac{c_s v^2/(2\Lambda)}{\mlb + c_s v^2 / (2\Lambda)} 
\; \approx \; \theta^2.
\end{eqnarray}
Note that here, the $\Lambda/m_n$ is replaced with $\Lambda/m_l = \Lambda/\Lambda$ which cancels.

Putting everything together, we have in the HSDM case
\begin{equation}
g_{B_d,h} \, = \, \frac{\mdm}{v} \theta^2 \left[ f_n^{(B_d)} \frac{\Lambda}{m_n} + \frac{4}{11N_D - 2} (1 - f_n^{(B_d)}) \right].
\end{equation}

\section{ALP coupling derivation \label{app:alp_coupling}}

Here we derive how \cref{eq:Lint_eta_fermions} is obtained from \cref{eq:Lint_eta_Higgs}.  Focusing on just the operator, our starting point is
\begin{equation}
\mathcal{O}_{\eta'} \, = \, \partial_\mu \eta'_d (H^\dagger i D^\mu H - (D^\mu H)^\dagger i H) \, ,
\label{eq:dim6cpoperator}
\end{equation}
where to be careful we have written out the Hermitian conjugate explicitly.  Integrating by parts, this becomes
\begin{eqnarray}
\mathcal{O}_{\eta'} &=& \eta'_d \big[-\partial_\mu H^\dagger i D^\mu H - H^\dagger i \partial_\mu (D^\mu H) \nonumber \\
& &{} + \partial_\mu (D^\mu H)^\dagger iH + (D^\mu H)^\dagger i \partial_\mu H \big] \, .
\end{eqnarray}
To simplify further, we can use the equation of motion for the Higgs field.  The Higgs part of the Lagrangian for the SM is as follows:
\begin{eqnarray}
\mathcal{L}_H &=& (D_\mu H)^\dagger D^\mu H + \mu^2 H^\dagger H - \lambda (H^\dagger H)^2 \nonumber \\
& &{} -y_e (\bar{L}_e H e_R + \bar{e}_R H^\dagger L_e)
\end{eqnarray}
where we include only the electron Yukawa coupling - other Yukawas are similar, and we will end up in the mass basis anyway.  Now we obtain the equation of motion:
\begin{eqnarray}
\frac{\delta \mathcal{L}_H}{\delta H^\dagger} &=& \partial_\mu \frac{\delta \mathcal{L}_H}{\delta (\partial_\mu H^\dagger)} \\
\mu^2 H - 2\lambda (H^\dagger H) H - y_e  \bar{e}_R L_e &=& \partial_\mu (D^\mu H) \, .
\end{eqnarray}
We have exactly the right-hand side in the operator above after integration by parts, but we also have the Hermitian conjugate.  Writing down the other EOM:
\begin{equation}
\mu^2 H^\dagger - 2\lambda H^\dagger (H^\dagger H) - y_e \bar{L}_e e_R \; = \; \partial_\mu (D^\mu H)^\dagger \, .
\end{equation}
Substituting in carefully, we have the following:
\begin{eqnarray}
\mathcal{O}_{\eta'} &=& -i\eta_d' \big[ \partial_\mu H^\dagger D^\mu H + \mu^2 H^\dagger H- 2\lambda (H^\dagger H)^2 \nonumber \\
& &{} - y_e \bar{e}_R H^\dagger L_e - \mu^2 H^\dagger H + 2\lambda (H^\dagger H)^2  \nonumber \\
& &{} + y_e \bar{L}_e H e_R - (D^\mu H)^\dagger \partial_\mu H \big] \nonumber \\
&=& -i\eta'_d \big[ \partial_\mu H^\dagger D^\mu H - (D^\mu H)^\dagger \partial_\mu H 
\nonumber \\
& &{} + y_e (\bar{L}_e H e_R - \bar{e}_R H^\dagger L_e) \big] \, .
\end{eqnarray}
Let's inspect the covariant derivative terms further.  The covariant derivative can be expanded out to give
\begin{equation}
D_\mu H \, = \, \partial_\mu H - igW_\mu^a \tau^a H  - \frac{ig'}{2} B_\mu H \, .
\end{equation}
To proceed from this point, we substitute in the Higgs field $H$ as $H \rightarrow (h + v) / \sqrt{2}$.  This comes with a selection of the lower SU$(2)$ index, which means that this substitution does the following:
\begin{eqnarray}
\bar{L_e} H e_R - \bar{e}_r H^\dagger L_e &\rightarrow& \frac{h+v}{\sqrt{2}} (\bar{e}_L e_R - \bar{e}_R e_L) \nonumber \\
&=& \frac{h+v}{\sqrt{2}} \bar{e} \gamma_5 e \, ,
\end{eqnarray}
and
\begin{eqnarray}
D_\mu H &\rightarrow& \frac{1}{\sqrt{2}} \left[ \partial_\mu h + \frac{ig}{2} (h+v) W_\mu^3 - \frac{ig'}{2} (h+v) B_\mu \right] 
\nonumber \\
&=& \frac{1}{\sqrt{2}} \left[ \partial_\mu h + i g_Z (h + v) Z_\mu \right] \, .
\end{eqnarray}
where $g_Z \equiv g/(2 \cos \theta_W)$. 
This leads to the following:
\begin{eqnarray}
\partial_\mu H^\dagger D^\mu H &\rightarrow& \frac{1}{2} \left[ \partial_\mu h \partial^\mu h + i g_Z (h+v) Z^\mu \partial_\mu h \right]. \qquad
\label{eq:differencecancel}
\end{eqnarray}
Now, when we take the difference $\partial_\mu H^\dagger D^\mu H - (D^\mu H)^\dagger \partial_\mu H$, the $(\partial_\mu h)^2$ terms will cancel out. 

In \cref{eq:differencecancel}, there is one term that seemingly remains, namely $\eta'_d \partial_\mu h Z^\mu$.  Explicitly, we can see this directly
from \cref{eq:dim6cpoperator} as
\begin{eqnarray}
c_6' \mathcal{O}_{\eta'} &\supset& 2 c_6' g_Z v h Z^\mu \partial_\mu \eta_d' \, .
\end{eqnarray}
However, this term does not actually contribute a physical interaction, which we now explain in detail.  First recognize that \cref{eq:dim6cpoperator}
also leads to $\eta'_d$--$Z^\mu$ mixing,
\begin{eqnarray}
    c_6' \mathcal{O}_{\eta'} &\supset& 2 c_6' g_Z v^2 Z^\mu \partial_\mu \eta_d' \, .
\label{eq:Zhetap}
\end{eqnarray}
This mixing is exactly how the $Z$ gauge boson
absorbs the Goldstone mode to acquire a mass.
Specifically, the normal
kinetic term for the Higgs boson is
\begin{eqnarray}
    - (D_\mu H)^\dagger D^\mu H &\supset& g_Z v Z^\mu \partial_\mu z'  \, ,
\label{eq:normalkineticterm}
\end{eqnarray}
where $z'$ is the would-be Goldstone boson absorbed
in $Z^\mu$ \emph{in the absence of} $\mathcal{O}_{\eta'}$.  In the presence of $c_6'$, however, 
these terms imply that the longitudinal mode $z$ absorbed by $Z^\mu$ is actually a linear combination of $z'$ and $\eta'$, 
\begin{eqnarray}
    z &=& z' + 2 c_6' v \eta'
\end{eqnarray}
Substituting for $z'$ back into the SM 
kinetic term \cref{eq:normalkineticterm}
gives the correct Goldstone mixing with $Z^\mu$
(by construction) along with an additional contribution
\begin{eqnarray}
    - (D_\mu H)^\dagger D^\mu H &\supset& 
    2 c_6' g_Z \eta' Z^\mu \partial_\mu h  \, .
\label{eq:normalkinetictermshift}
\end{eqnarray}
The term \cref{eq:normalkinetictermshift}
has the same sign as \cref{eq:Zhetap},
and so they can be combined, up to a total derivative term in the Lagrangian, to give
\begin{equation}
    2 c_6' g_Z v \left( h Z^\mu \partial_\mu \eta_d' +
    \eta' Z^\mu \partial_\mu h \right) 
    \; \rightarrow \; 
    2 c_6' g_Z v h \eta' (\partial_\mu Z^\mu) \, .
\label{eq:normalkinetictermshift2}
\end{equation}
This combined term is gauge-redundant, and can be removed 
by a suitable modification of the 
gauge-fixing terms, thereby not contributing
to the action.

Finally, we see that the only coupling that remains is precisely
\begin{eqnarray}
\mathcal{O}_{\eta'} &\supset& -i\eta'_d \frac{y_e}{\sqrt{2}} (h+v) \bar{e} \gamma_5 e \nonumber \\
&=&{} -\frac{m_e}{\sqrt{2}} \eta_d' \left(1 + \frac{h}{v} \right) \bar{e} i\gamma_5 e
\end{eqnarray}
which is exactly the result from \cite{Bauer:2016zfj} and what we used to obtain the fermion bilinear coupling above in \cref{eq:Lint_eta_fermions}.

\raggedright
%


\end{document}